\documentclass[a4paper,manyauthors,nocleardouble,COMPASS]{cernphprep}
\usepackage[a4paper, total={17 cm, 20cm}]{geometry}

\RequirePackage[T1]{fontenc}

\RequirePackage{graphicx}
\RequirePackage{newtxtext}
\RequirePackage{newtxmath}
\RequirePackage{flushend}
\usepackage{tikz}
\usepackage{lineno}
\usepackage[normalem]{ulem} 
\usepackage{soul} 
\usepackage{amsmath}
\usepackage{balance}
\usepackage{lastpage}
\usepackage{caption}
\usepackage{subcaption}
\usepackage{multirow}
\usepackage{bigstrut}
\usepackage{widetext}

\usepackage{orcidlink}
\usepackage{tikz}
\usepackage{hyperref}
\hypersetup{
    colorlinks,
    citecolor=blue,
    filecolor=blue,
    linkcolor=blue,
    urlcolor=blue
}

\def\ptv{\textbf{\textit{P}}_{\rm{T}}}
\def\absptv{P_{\rm{T}}}

\def\pt{P_{\rm{T}}}
\def\pperp{\textbf{\textit{p}}_{\rm \perp}}
\def\STv{\textbf{\textit{S}}_{\rm{T}}}
\def\ptvptv{P^2_{\rm{T}}}
\def\gevcc{\rm{GeV}/\textit{c}}
\def\gevc2{\rm{GeV}/\textit{c}^2}
\def\Mhh{\textit{M}_{\rm{hh}}}
\def\pht{\textit{P}_{\rm{T}}}
\def\pt{\textbf{\textit{p}}_{\rm \perp}}

\def\kt{\textbf{\textit{k}}_{\rm T}}

\def\phih{\phi_{\rm{hh}}}
\def\phis{\phi_{\rm{S}}}
\def\phiC{\phi_{\rm{Coll}}}
\def\phiSiv{\phi_{\rm{Siv}}}
\def\Dnn{D_{\rm{NN}}}

\def\z{z}
\def\rhoz{$\rho^0$}
\def\aUTX{a_{\rm{UT}}^{\sin\phi_{\rm X}}}
\def\aUTC{a_{\rm{UT}}^{\sin\phi_{\rm Coll}}}
\def\aUTS{a_{\rm{UT}}^{\sin\phi_{\rm Siv}}}
\def\AUTbgX{A_{\rm{UT},bg}^{\sin\phi_{\rm X}}}
\def\AUTX{A_{\rm{UT}}^{\sin\phi_{\rm X}}}
\def\xu{\hat{\textbf{x}}}
\def\zu{\hat{\textbf{z}}}
\def\VM{\rm{VM}}

\begin{document}
\begin{titlepage}
\PHnumber{2022-232}
\PHdate{\today}
\title{Collins and Sivers transverse-spin asymmetries in inclusive muoproduction of $\rho^0$ mesons}

\Collaboration{The COMPASS Collaboration}
\ShortAuthor{The COMPASS Collaboration}

\begin{abstract}
The production of vector mesons in deep inelastic scattering is an interesting yet scarcely explored channel to study the transverse spin structure of the nucleon and the spin-dependence of fragmentation. The COMPASS collaboration has performed the first measurement of the Collins and Sivers asymmetries for inclusively produced $\rho^0$ mesons. The analysis is based on the data set collected in deep inelastic scattering in $2010$ using a $160\,\,\rm{GeV}/c$ $\mu^+$ beam impinging on a transversely polarized $\rm{NH}_3$ target. The $\rho^{0}$ mesons are selected from oppositely charged hadron pairs, and the asymmetries are extracted as a function of the Bjorken-$x$ variable, the transverse momentum of the pair and the fraction of the energy $z$ carried by the pair. Indications for positive Collins and Sivers asymmetries are observed.
\end{abstract}

\vfill
\Submitted{(accepted at Phys. Lett. B)}

\end{titlepage}
{
\pagestyle{empty}
\clearpage
}
\clearpage

\setcounter{page}{1}

\section{Introduction}\label{sec:intro}
In semi-inclusive deep inelastic scattering (SIDIS) $l\,N\rightarrow l'\,h\,X$, a high-energy lepton $l$ scatters off a target nucleon $N$, and in the final state the scattered lepton $l'$ is observed in coincidence with at least one hadron $h$ produced in the current fragmentation region.
This process presently is the main tool to study the 3-dimensional partonic structure of the nucleon, i.e. the transverse spin and transverse momentum distributions of partons, and the possible correlations between their spin, their motion, and the spin of the nucleon. In the present quantum chromodynamics (QCD) framework, such information is encoded in the Transverse Momentum Dependent Parton Distribution Functions (TMD PDFs).
In SIDIS, the above mentioned correlations induce azimuthal asymmetries in the angular distributions of the produced hadrons, which are interpreted in terms of convolutions of TMD PDFs and Transverse Momentum Dependent Fragmentation Functions (TMD FFs) \footnote{For TMD PDFs and TMD FFs we use the notation of Ref. \cite{Mulders:1995dh}.}.

Among the accessible asymmetries, the transverse single spin asymmetries (TSAs), which arise for a transversely polarized target nucleon, have been extensively studied in recent years.
In particular, the Collins asymmetry \cite{collins} arises from the convolution between the chiral-odd transversity PDF $h_1^q$ \cite{Jaffe-Ji-transversity-1990} and the chiral-odd and T-odd Collins FF $H_{1q}^{\perp\,h}$ \cite{collins}. The transversity PDF, which is the difference between the number density of partons with transverse polarisation parallel and antiparallel to the transverse polarisation of the parent nucleon, is the least well known among the three collinear PDFs needed for the complete characterization of the nucleon structure at leading order. The Collins TMD FF describes the correlation between the transverse polarisation of a fragmenting quark and the transverse momentum of the produced hadron, and probes the quark-spin dependence of the fragmentation process.

Another important TSA is the so-called Sivers asymmetry, which is interpreted as the convolution of the Sivers function $f_{1 \rm T}^{\perp\,q}$ \cite{Sivers1990}, the TMD PDF that describes the transverse momentum distribution of an unpolarized quark in a transversely polarized nucleon, and the spin-averaged TMD FF $D_{1q}^h$ that describes the fragmentation of an unpolarized quark into an unpolarized hadron.

Collins and Sivers asymmetries have been measured since 2005, in particular in SIDIS off protons, deuterons or neutrons for unidentified charged hadrons and for identified pions, kaons and protons by the HERMES \cite{HERMES:2020}, COMPASS \cite{COMPASS:2012coll,COMPASS:2012siv,COMPASS-collins-sivers} and JLab \cite{jlab-ssa} experiments. Phenomenological analyses of the Collins asymmetries and the corresponding asymmetries measured in $e^+e^-$ annihilation to hadrons \cite{belle-spin-asymmetries,BESIII-Collins,BABAR-Collins} have led to the extraction of both the transversity PDF and the Collins FF \cite{Anselmino-belle-2015,M.B.B}. Similarly, from the HERMES and COMPASS measurements of the Sivers asymmetry, the Sivers function was extracted by several authors \cite{Anselmino:2012aa,Martin:2017yms,COMPASS:2018ofp}. For a recent review on transverse spin effects on semi-inclusive processes see, \textit{e.g.} Ref. \cite{Anselmino:2020vlp}.

Relevant information on nucleon structure and the fragmentation process can also be obtained from measurements of TSAs in inclusive production of vector mesons in DIS \cite{Bacchetta:Spin1}. The Collins asymmetry for vector mesons couples the transversity PDF to the Collins FF $H_{1q}^{\perp \rm{VM}}$, which describes the fragmentation of a transversely polarized quark into a vector meson. The investigation of this channel could shed new light on the still poorly known quark-spin dependence of the fragmentation process. Model predictions based on the recursive string+${}^3P_0$ model of polarized quark fragmentation \cite{Czyzewski-vm,Kerbizi:2021} suggest that the Collins asymmetry for $\rho$ mesons has the opposite sign compared to that of positive pions. Depending on the choice of the parameters, the Collins asymmetry for vector meson production can be as large as for positive pions.

Up to now, transverse spin asymmetries for vector meson production have not been measured, neither in SIDIS nor in $e^+e^-$ annihilation into hadrons.
The low statistics of the produced vector mesons and the high combinatorial background make the measurement of these asymmetries very challenging.

In this article we present the first measurement of the Collins and Sivers asymmetries for \rhoz{} mesons produced in DIS off a transversely polarized proton target. The analysis is performed on the COMPASS data collected in 2010. The same data have already been used for many published results, e.g., the Collins and Sivers asymmetries for unidentified charged hadrons \cite{COMPASS:2012coll,COMPASS:2012siv}, pions and kaons \cite{COMPASS-collins-sivers}, and dihadron production asymmetries \cite{compass-dihadron}. The final data set used for the analysis described in this paper consists of $2.6\times 10^6$ $\rho^0$ mesons.

The article is organized as follows. The formalism of vector meson production in SIDIS is introduced in Sec. \ref{sec:formalism}. Section \ref{sec:setup} describes the experimental apparatus and the data set used for this analysis. The method used for the extraction of TSAs is explained in Sec. \ref{sec:procedure}. The extraction of the \rhoz{} signal is described in Sec. \ref{sec:signal} and the results for Collins and Sivers \rhoz{} asymmetries are given in Sec. \ref{sec:asymmetries}. In Sec. \ref{sec:conclusions} conclusions are drawn.

\section{Theoretical formalism}\label{sec:formalism}
\begin{figure}[h]
    \centering
     \begin{minipage}[b]{0.4\textwidth}
         \centering
         \includegraphics[width=\textwidth]{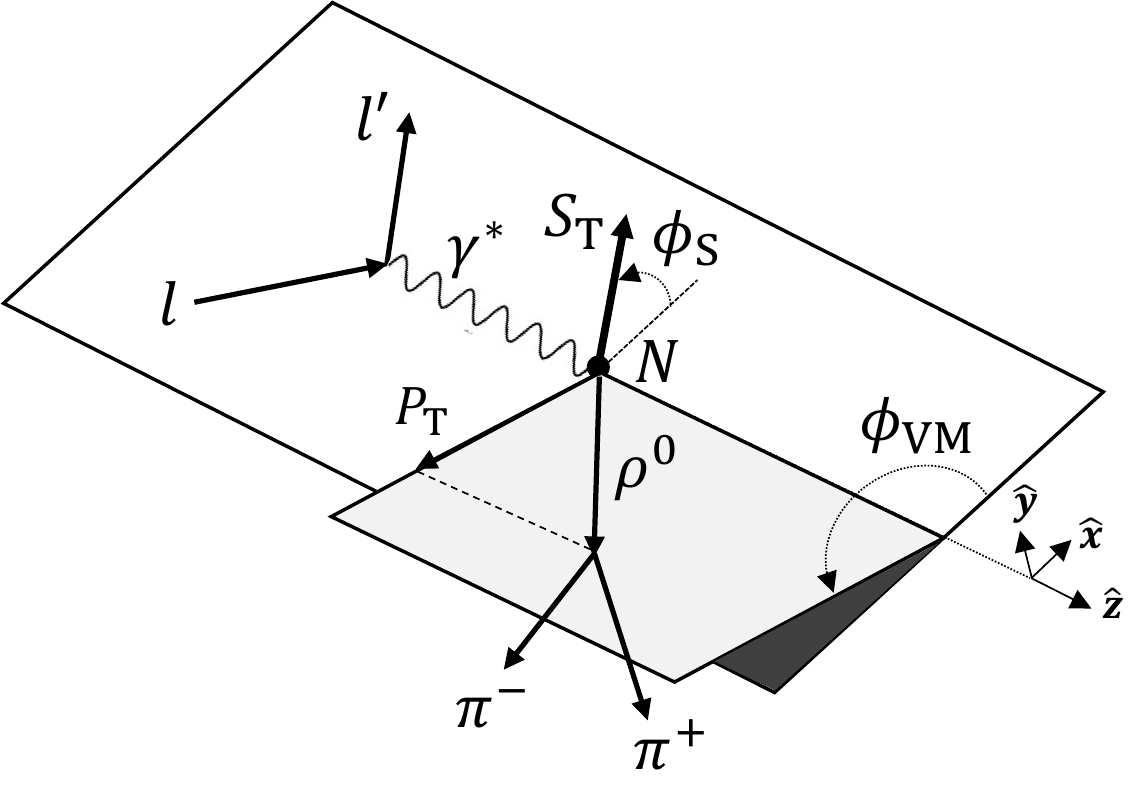}
     \end{minipage}
        \caption{Kinematics of the SIDIS process for \rhoz{} meson production in the gamma-nucleon reference system.}
        \label{fig:SIDIS kinematics}
\end{figure}

The kinematics for the production of a vector meson ($\VM$) in SIDIS off transversely polarised protons in the one-photon exchange approximation is schematically shown in Fig. \ref{fig:SIDIS kinematics}.
The process is represented in the gamma-nucleon system (GNS), namely in the reference system where the momentum of the exchanged virtual photon defines the $\zu$ axis and the $\xu$--$\zu$ plane is the lepton scattering plane with the $\xu$ axis along the transverse component of the lepton momenta \cite{Bacchetta:2004jz}.

If the polarisation of the vector meson is not considered, at leading twist TMD PDFs and TMD FFs and leading order order QCD, the differential cross section of the process $l\,N\rightarrow l'\,\rho^0\,X$ is \cite{Bacchetta-SIDIS}:
\begin{widetext}
\begin{eqnarray}\label{eq:cross section}
\nonumber \hspace{-1cm}\frac{d^6\sigma}{dx\,dy\,dz\,d\ptvptv\,d\phi_{\VM}\,d\phis}&=&\frac{\alpha^2}{x\,y\,Q^2}\,\left(\frac{1+(1-y)^2}{2}\right)\, \times\\ \nonumber
 &\times& \Bigg\{ \sum_{q} e_q^2\mathcal{C}\,\left[f_1^q\,D_{1q}^{\VM}\right]
+\Dnn\,|\STv|\,\sum_{q} e_q^2\mathcal{C}\,\left[\frac{\pperp\cdot\ptv}{z\,M_{\rm{VM}}\,P_{\rm T}}\,h_1^q\,H_{1q}^{\perp \VM}\right]\, \times\\
 &\times& \sin\left(\phi_{\rm VM}+\phis-\pi\right) + |\STv|\,\sum_{q} e_q^2\mathcal{C}\,\left[\frac{\kt\cdot\ptv}{M\, P_{\rm T}}\,f_{1\rm T}^{\perp q}\,D_{1q}^{\VM}\right]\,\sin\left(\phi_{\rm VM}-\phis\right) + \dots \Bigg\}.
\end{eqnarray}
\end{widetext}
Here, $x$ is the Bjorken variable, $y$ is the fraction of the initial lepton energy loss in the target rest frame, and $Q^2$ is the virtuality of the photon. The variable $z$ is the fraction of the energy of the virtual photon carried by the produced VM in the target rest frame and $\absptv$ is the modulus of its transverse momentum $\ptv$ in the GNS.
The variables $\phi_{\rm VM}$ and $\phis$ are the azimuthal angle of $\ptv$ and the one of the target transverse polarisation $\STv$ in the GNS, respectively. The combinations of angles $\phiC=\phi_{\VM}+\phis-\pi$ and $\phiSiv=\phi_{\VM}-\phis$ are the Collins angle and the Sivers angle associated to the transverse momentum of the VM, respectively. The factor $\Dnn=(1-y-\gamma^2y^2/4)/(1-y+y^2/2+\gamma^2y^2/4)$, where $\gamma=2Mx/Q$, is the virtual photon depolarisation factor. The nucleon mass and the mass of the vector meson are denoted as $M$ and $M_{\VM}$, respectively. The summations run over the quark and antiquark flavours and the (anti)quark charge $e_q$ is given in units of the elementary charge.
The cross section in Eq.~(\ref{eq:cross section}) is written in terms of structure functions each involving the convolution
\begin{align}
\nonumber \mathcal{C}\left[wfD\right]&=\int d^2\kt\, d^2\pt\,\delta^{(2)}(z\kt+\pt-\ptv) \\
&\times w(\kt,\pt)\,\, xf(x,\kt)\,\, D(z,\pt),
\end{align}
 where $f$ indicates a TMD PDF, $D$ indicates a TMD FF, and $w$ is a weight factor depending on the intrinsic quark transverse momentum $\kt$ in the GNS and on the transverse momentum $\pt$ of the VM with respect to the direction of the scattered quark.

 The expression for the Collins asymmetry can be obtained from the ratio between the transverse-spin-dependent and the spin-averaged terms of the cross section in Eq.~(\ref{eq:cross section}), and is given by
 \begin{eqnarray}\label{eq:Collins parton model}
 A_{\rm{UT}}^{\sin\phiC}(x,z,\absptv)=\frac{\sum_{q} e_q^2\mathcal{C}\,\left[\frac{\pperp\cdot\ptv}{z\,M_{\VM}\,P_{\rm T}}\,h_1^q\,H_{1q}^{\perp \VM}\right]}{\sum_{q} e_q^2\mathcal{C}\,\left[f_1^q\,D_{1q}^{\VM}\right]}.
 \end{eqnarray}
 The functions $D_{1q}^{\VM}$ and $H_{1q}^{\perp \VM}$ describe the fragmentation of an unpolarized and a transversely polarized quark $q$ into a vector meson, respectively.

The Sivers asymmetry reads \cite{Bacchetta-SIDIS}
 \begin{eqnarray}\label{eq:Sivers parton model}
 A_{\rm{UT}}^{\sin\phiSiv}(x,z,\absptv)=\frac{\sum_{q} e_q^2\mathcal{C}\,\left[\frac{\kt\cdot\ptv}{M\,P_{\rm T}}\,f_{1 \rm T}^{\perp q}\,D_{1q}^{\VM}\right]}{\sum_{q} e_q^2\mathcal{C}\,\left[f_1^q\,D_{1q}^{\VM}\right]}.
 \end{eqnarray}

The expected number of vector mesons is
\begin{align}\label{eq: N_h1h2}
\nonumber N_{\rm{VM}}(x,\z,\absptv,\phiC,\phiSiv)\propto\big(1&+\Dnn\,f\,P_{\rm t}\,A_{\rm{UT}}^{\sin\phiC}\,\sin\phiC \\
&+ f\,P_{\rm t}\,A_{\rm{UT}}^{\sin\phiSiv}\,\sin\phiSiv\big),
\end{align}
where the other possible modulations are neglected. Here, $f$ is the dilution factor that takes into account the fraction of polarisable protons in the target, and $P_{\rm t}$ is the average transverse polarisation of the polarisable protons in the target.

For the specific case of $\rho^0$ meson production considered in this work, we use all oppositely charged hadron pairs in the event. The four-momentum of the \rhoz{} candidate is given by $P_{\rm{h}_1}+P_{\rm{h}_2}$, where $P_{\rm{h}_1}$ and $P_{\rm{h}_2}$ are the momenta of the positive and the negative hadron of the pair, respectively. The fractional energy of the \rhoz{} candidate is given by $z=z_{\rm{h}_1}+z_{\rm{h}_2}$ and its transverse momentum in the GNS is given by $\ptv=\textbf{\textit{P}}_{\rm{h}_1\rm{T}}+\textbf{\textit{P}}_{\rm{h}_2\rm{T}}$, where we have indicated by $z_{\rm{h}_i}$ and $\textbf{\textit{P}}_{\rm{h}_i\rm T}$ ($i=1,2$) the fraction of the virtual photon energy carried by the hadron $h_{i}$ in the target rest system, and the transverse momentum of $h_i$ in the GNS, respectively. In the following we indicate with $\phih$ the azimuthal angle of $\ptv$, and with $M_{\rm{hh}}=\sqrt{(P_{\rm{h}_1}+P_{\rm{h}_2})^2}$ the invariant mass of the pair. As it will be discussed in Sec. \ref{fig:sig fraction}, the data set of \rhoz{} candidates contains a sizeable combinatorial background due to non-resonant hadron pairs.

\begin{figure*}[tb]
     \centering
     \begin{subfigure}[b]{0.30\textwidth}
         \centering
         \includegraphics[width=\textwidth]{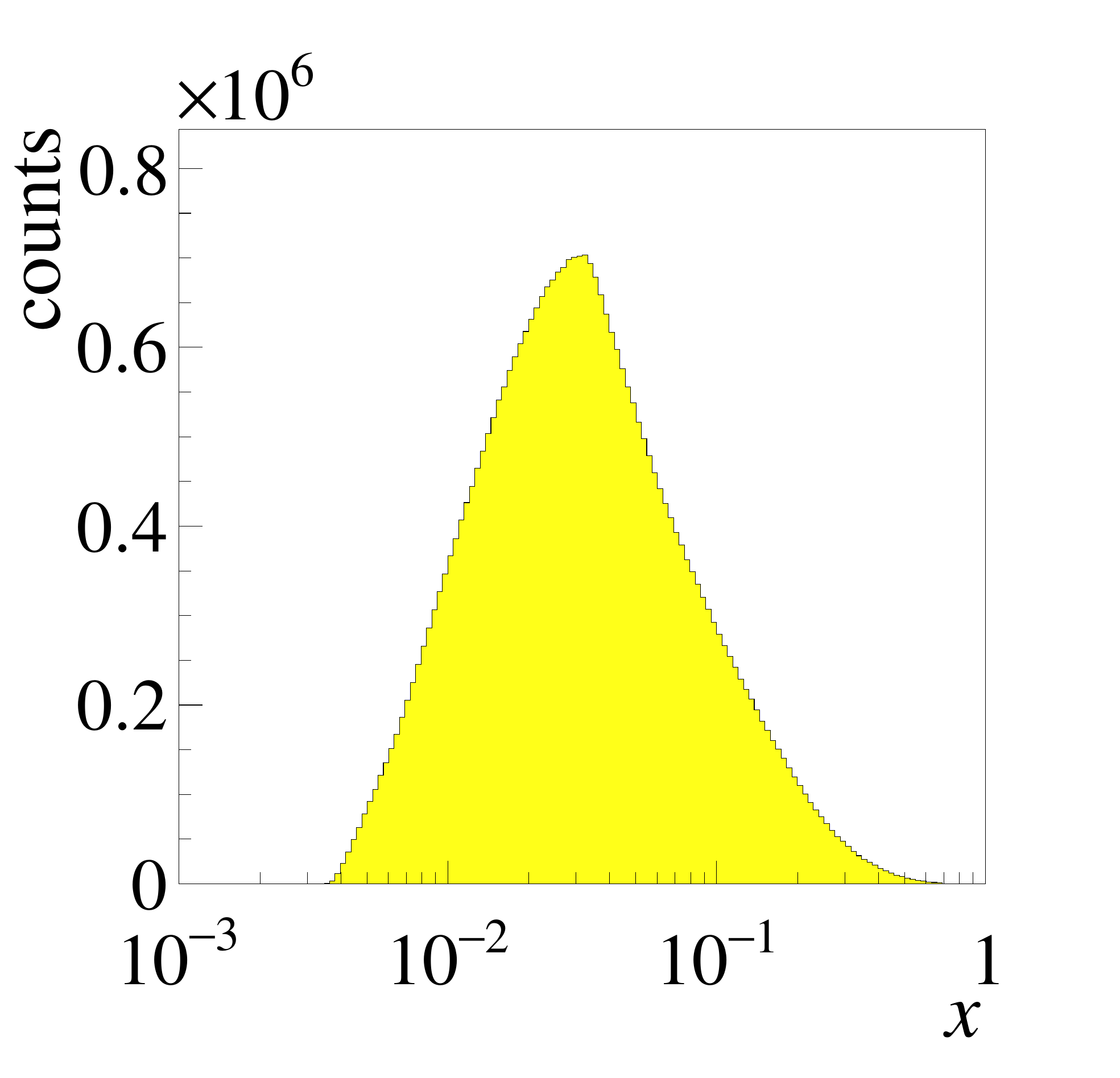}
     \end{subfigure}
     \begin{subfigure}[b]{0.30\textwidth}
         \centering
         \includegraphics[width=\textwidth]{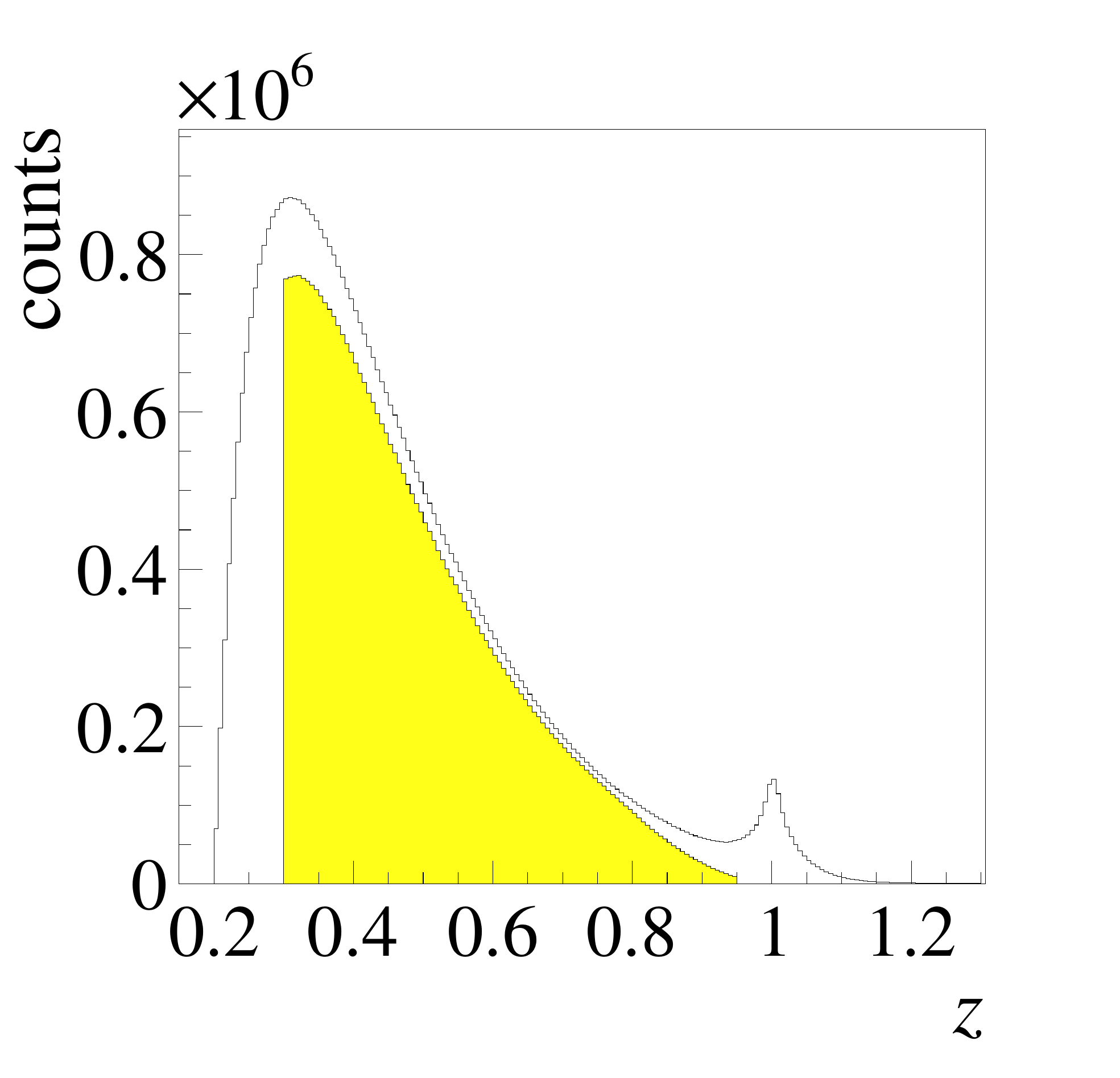}
     \end{subfigure}
     \begin{subfigure}[b]{0.30\textwidth}
         \centering
         \includegraphics[width=\textwidth]{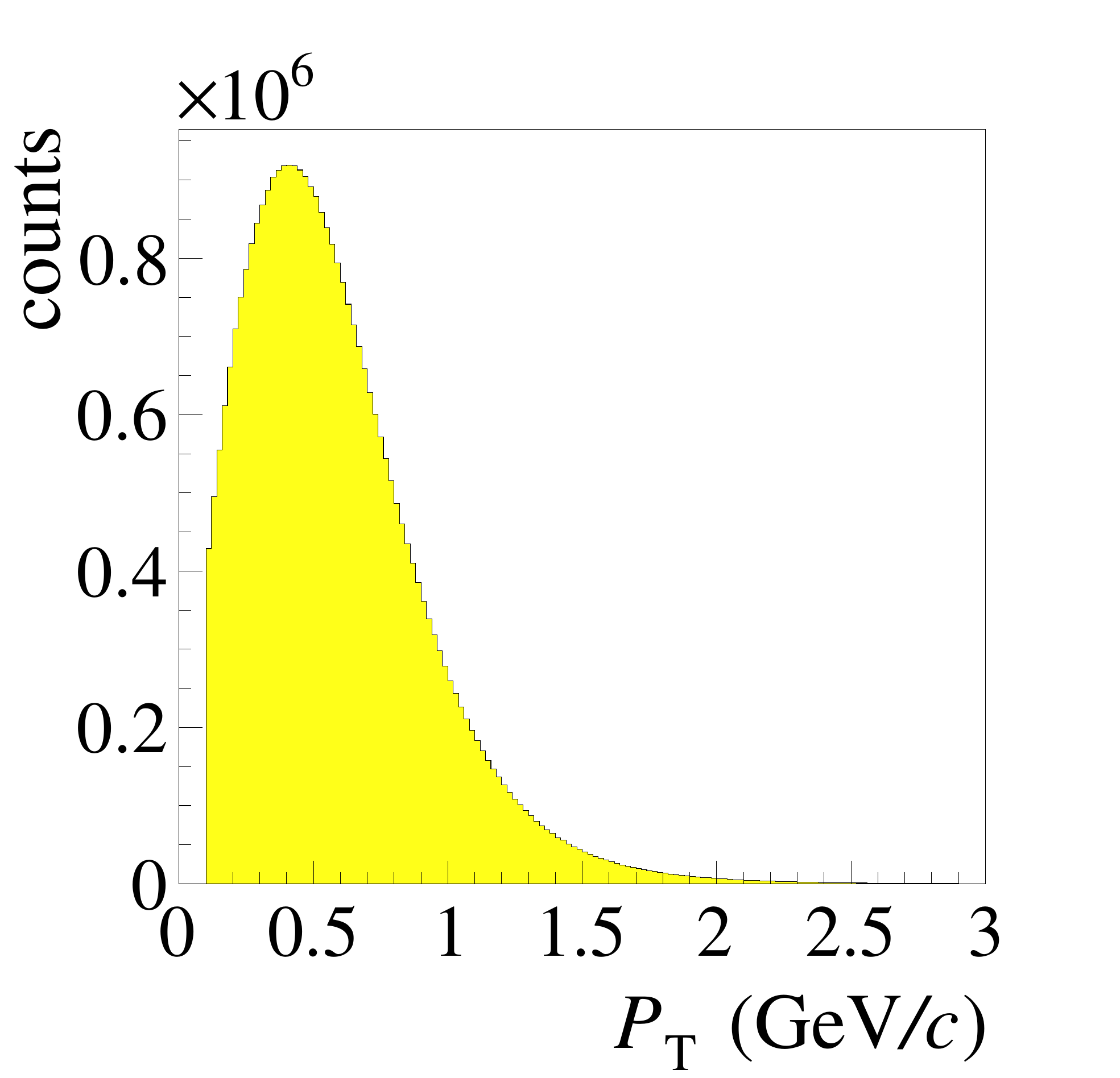}
     \end{subfigure}
        \caption{Distributions of $x$ (left panel), $z$ (middle panel) and $\absptv$ (right panel) for the selected $h^+h^-$ pairs (filled histograms). The empty histogram in the middle panel shows the $z$ distribution before applying the requirements on hadron pairs.}
        \label{fig:xzPT}
\end{figure*}

\section{Experimental apparatus and data sample}\label{sec:setup}
The COMPASS experiment, a fixed target experiment located at the M2 beamline of the CERN SPS, is in operation since 2002. A detailed description of the apparatus can be found in Ref. \cite{COMPASS:2007rjf,COMPASS:2012coll,COMPASS-collins-sivers}. The data used in this analysis were collected in 2010 using a $160\,\rm{GeV}/c$ $\mu^+$ beam and a transversely polarized $\rm{NH}_{3}$ target. The target consisted of three cylindrical cells with neighbouring cells polarized in opposite directions in order to collect data simultaneously for both target spin orientations. The average polarisation of the hydrogen nuclei was $\langle P_{\rm t}\rangle\simeq 0.8$. The average dilution factor was $\langle f\rangle\simeq 0.15$, with a slight dependence on $x$ and constant in $z$ and $P_{\rm hT}$ \cite{COMPASS:2012siv}. The data taking was divided in twelve periods of about ten days. In order to compensate for acceptance effects the polarisation was reversed in the middle of each period.

Only events with an incoming and an outgoing muon track and at least two produced charged hadrons are considered. Equal flux along the target cells is assured by requiring the extrapolated incoming muon to cross all three target cells. We only consider hadrons coming from the muon production vertex, thus tracks from mesons produced in weak decays do not enter this analysis. In order to ensure the deep inelastic regime, we require $Q^2 > 1.0\, (\rm{GeV}/c)^2$ and the invariant mass of the final hadronic system $W > 5\, \rm{GeV}/c^2$. The Bjorken $x$ variable ranges between $0.003$ and $0.7$. The selection $y>0.1$ removes events with poorly reconstructed virtual photon energy and $y<0.9$ removes events with large radiative effects. For hadrons, we require $z_h>0.1$ to ensure the current fragmentation regime and $P_{h\rm T}>0.1 \,\rm{GeV}/c$ to ensure good resolution in the respective azimuthal angle.

The dihadron samples with pairs $h^+h^-, h^+h^+, h^-h^-$ of charged hadrons are selected as described in the following. In order to avoid the contribution from non-SIDIS diffractive events, exclusively produced $h^+h^-$ pairs are rejected by requiring a missing energy  $E_{\rm{miss}} > 3.0\,\rm{GeV}$, where $E_{\rm{miss}}=(M_{\rm X}^2-M^2)/2M$, and $M_{\rm X}^2=(q+P_{\rm p}-P)^2$ where $q$, $P_{\rm p}$ and $P$ are the four momenta of the exchanged photon, the target proton and the hadron pair respectively. The requirements $z<0.95$, $0.1\,\rm{GeV}/c < \pht < 4.0\,\rm{GeV}/c$ and $0.35\,\gevc2 < \Mhh < 3.0\,\gevc2$ are also applied in order to define the kinematic range of the analysis. The selection $z>0.3$ is applied in order to enhance the fraction of \rhoz{} mesons. No further selection improving the signal over background ratio could be found.

The selected dihadron samples consist of about $3.4\times 10^7$ $h^+h^-$ pairs, $1.1\times 10^7$ $h^+h^+$ pairs and $0.7\times 10^7$ $h^-h^-$ pairs. The distributions of $x$, $z$ and $P_{\rm T}$ for the selected $h^+h^-$ sample are shown as filled histograms in the left, middle and right panels of Fig. \ref{fig:xzPT}, respectively. The empty histogram in the middle panel of the same figure shows the $z$ distribution without the requirements on the hadron pairs. The exclusive peak at $z=1$, rejected by the requirement on the missing energy, is clearly visible.

\begin{figure}[tb]
     \centering
     \begin{subfigure}[b]{0.50\textwidth}
         \centering
         \includegraphics[width=\textwidth]{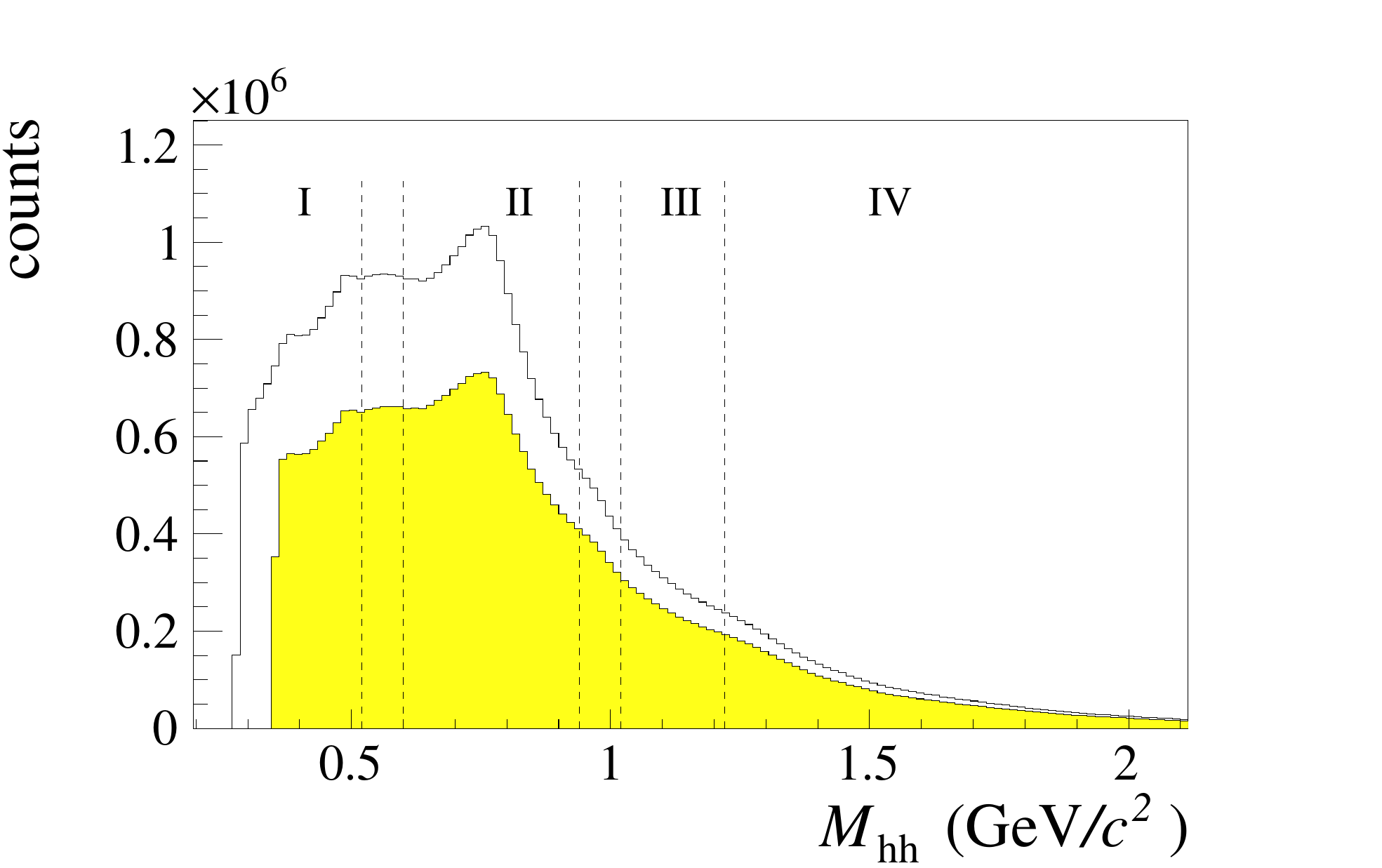}
     \end{subfigure}
        \caption{Invariant mass distribution of $h^+h^-$ pairs before (empty histogram) and after (filled histogram) applying the requirements on the hadron pair variables. The vertical lines indicate the different invariant mass regions defined in Tab. \ref{tab:regions}.}
        \label{fig:M}
\end{figure}

The invariant mass distribution for the $h^+h^-$ sample is shown in Fig. \ref{fig:M}, where the empty (filled) histogram is the invariant mass distribution before (after) the application of the cuts on the hadron pair variables. The peak corresponding to the $\rho^0(770)$ invariant mass is clearly visible as well as the broad structures corresponding to the $f_0(980)$ and $f_2(1240)$ mesons. As the requirement of the missing energy rejects exclusive events, it also reduces the significance of the \rhoz{} peak.

\section{Method for the extraction of the transverse single spin asymmetries}\label{sec:procedure}

The invariant mass range is divided in the four regions that are defined in Tab. \ref{tab:regions} and shown in Fig. \ref{fig:M} as separated by the vertical lines.
Region II covers the \rhoz{} invariant mass peak and will be referred to as the ``\rhoz{} region'' in the following. Regions I and III are dominated by the combinatorial background and in the following will be referred to as the ``side band'' regions. Region IV has a comparable number of pairs as region III and is included to study the invariant mass dependence of TSAs for the combinatorial background, although it is not used in the extraction of the $\rho^0$ asymmetries.

\begin{table}[t]
\centering
\begin{small}
 \begin{tabular}{|c| c|}
 \hline
 Region & Invariant mass range \\ [0.5ex]
 \hline\hline
 I\bigstrut[t] & $0.35\,\gevc2<\Mhh<0.52\,\gevc2$  \\
 II\bigstrut[t] & $0.60\,\gevc2<\Mhh<0.94\,\gevc2$  \\
 III\bigstrut[t] &  $1.02\,\gevc2<\Mhh<1.22\,\gevc2$ \\
 IV\bigstrut[tb] & $1.22\,\gevc2<\Mhh< 3.00\,\gevc2$ \\
 \hline
 \end{tabular}
 \caption{Invariant mass regions used in the extraction of the asymmetries.}
\label{tab:regions}
\end{small}
\end{table}

The extraction of the \rhoz{} TSAs proceeds using the following steps. First, the fraction $f_{s}$ of \rhoz{} mesons in the \rhoz{} region is evaluated. Then, the transverse single spin asymmetry $\aUTX$ of the $h^+h^-$ pairs in the \rhoz{} region is extracted. The angle $\phi_{\rm X}$ indicates either the Collins angle $\phiC$ or the Sivers angle $\phiSiv$. The asymmetries are measured for all kinematic bins by using the unbinned maximum likelihood method \cite{compass-dihadron} and the fit function $F(\phiC,\phiSiv)=a_{\rm 0}\times(1+\Dnn\,f\,P_{\rm t}\,a_{\rm C}\,\sin\phiC+f\,P_{\rm t}\,a_{\rm S}\,\sin\phiSiv)$ in agreement with the r.h.s. of Eq.~(\ref{eq: N_h1h2}). The parameters $a_{\rm 0}$, $a_{\rm C}$ and $a_{\rm S}$ are determined from the fitting procedure and are used to calculate the asymmetries $\aUTC=a_{\rm C}$ and $\aUTS=a_{\rm S}$. The transverse single spin asymmetry $\AUTbgX{}$ of the background is evaluated as the mean of the asymmetries in regions I and III using the same procedure. In order to obtain the asymmetry for \rhoz{} mesons, the background transverse single spin asymmetry is subtracted from the asymmetry in the \rhoz{} region according to
    \begin{equation}\label{eq:subtraction}
        \AUTX = \left[\aUTX-(1-f_s)\,\AUTbgX\right]\times \frac{1}{f_s}.
    \end{equation}

\section{Estimation of the $\rho^0$ signal}\label{sec:signal}
\begin{figure}[tb]
     \centering
     \begin{subfigure}[b]{0.5\textwidth}
         \centering
         \includegraphics[width=\textwidth]{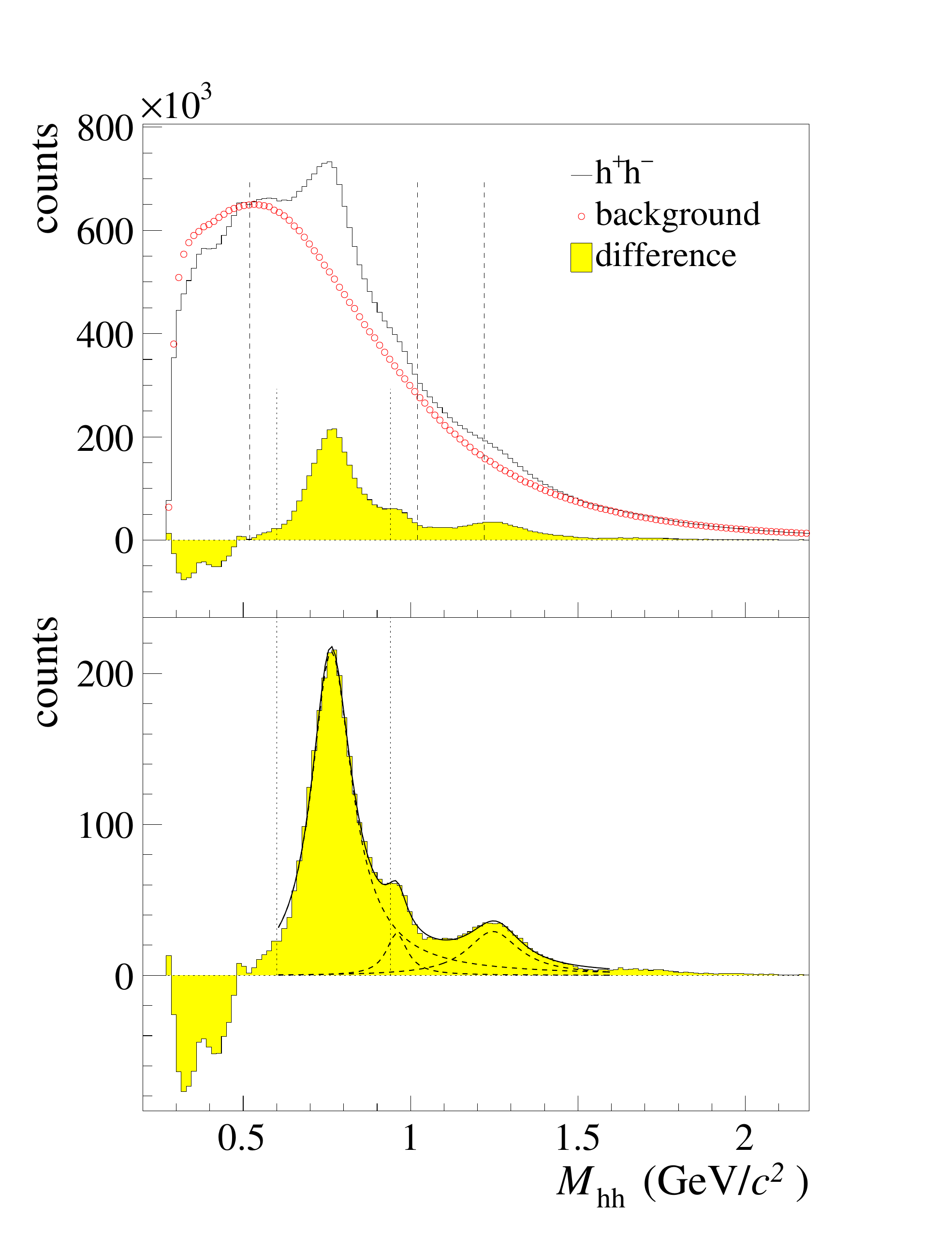}
     \end{subfigure}
        \caption{Top panel: the invariant mass distribution of $h^+h^-$ pairs (empty histogram), the background distribution (red points) and the difference between the two distributions (filled histogram). Bottom panel: zoom of the difference distributions. The vertical lines show the invariant mass regions defined in Tab. \ref{tab:regions}. See text for more details.}
        \label{fig:sig extraction}
\end{figure}

In order to determine the fraction of \rhoz{} mesons in the invariant mass region II, it is necessary to evaluate the contribution of the combinatorial background. The shape of the background distribution in the $\rho^0$ region is taken from the sum of the invariant mass distributions of $h^+h^+$ and $h^-h^-$ pairs. As normalisation the ratio between the number of $h^+h^-$ pairs and the number of like-sign pairs in the invariant mass interval $0.46\,\gevc2<\Mhh<0.56\,\gevc2$ is used, where the ratio between the invariant mass distributions of opposite and like-sign pairs is constant. For $\Mhh<0.50\,\gevc2$ the background subtraction procedure yields negative counts, meaning that the estimated background is larger than the $h^+h^-$ distribution. The distribution of like-sign pairs is in fact not expected to reproduce the full invariant mass spectrum of $h^+h^-$ pairs because of the different resonant structures of like-sign and unlike-sign pairs. However it can be used to estimate the background in the \rhoz{} region. The estimated invariant mass distribution of the background is shown in the top panel of Fig. \ref{fig:sig extraction} by the red points. It is subtracted from the $h^+h^-$ distribution (continuous histogram) to obtain the difference of the distributions shown by the filled histogram. After checking the compatibility of the invariant mass distributions in the different data taking periods, this procedure is performed on the invariant mass distributions integrated over the full year of data taking.

\begin{figure}[tb]
     \hspace{-3.0em}
     \centering
     \begin{subfigure}[b]{0.7\textwidth}
         \centering
         \includegraphics[width=1.0\textwidth]{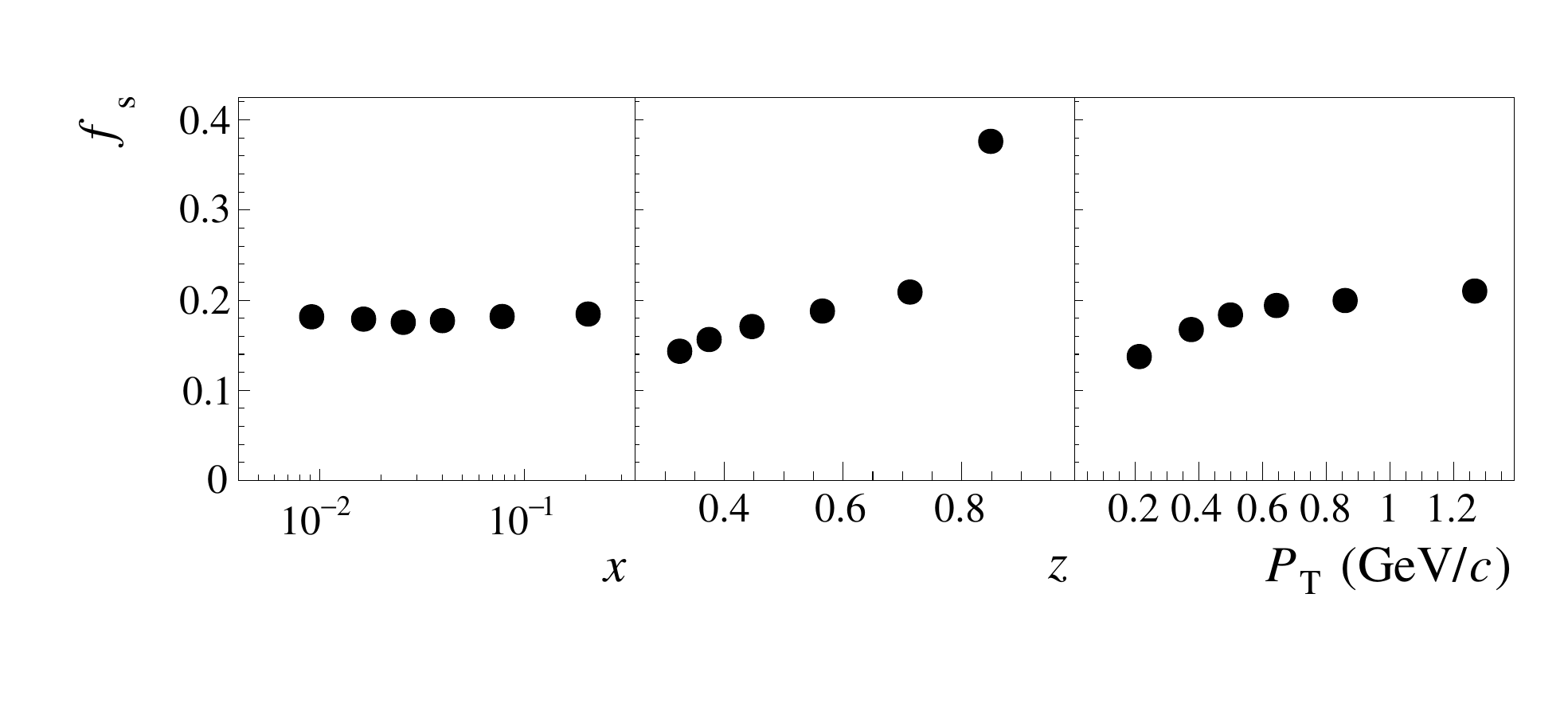}
     \end{subfigure}
     \vspace{-2.0em}
    \caption{The \rhoz{} signal fraction as a function of $x$ (left), $z$ (middle) and $\absptv$ (right).}
    \label{fig:sig fraction}
\end{figure}

In the difference of distributions, shown more clearly in the bottom panel of Fig. \ref{fig:sig extraction}, the peak corresponding to the $\rho^0(770)$ meson is clearly visible. Also visible are the structures corresponding to the $f_0(980)$ and $f_2(1270)$ mesons.
The difference of the distributions is fitted successfully by a sum of three Breit-Wigner functions\footnote{A $p$-wave Breit-Wigner function is used to describe the $\rho^0$ peak and a $s$-wave Breit-Wigner function is employed to describe the $f_0$ and $f_2$ peaks.}, demonstrating that the subtraction procedure is clean. In each Breit-Wigner function the parameters corresponding to the nominal mass and the width of the resonance are fixed to the corresponding PDG values, and only the normalisation parameter is estimated by the fit procedure. The fit function is shown by the continuous line, and the separate contributions of the $\rho^0$, $f_0$ and $f_2$ mesons are shown by the dashed lines. The fact that the extracted \rhoz{} distribution can be successfully described by a Breit-Wigner function provides confirmation that the form of the combinatorial background in region II and its normalisation is evaluated correctly.

\begin{figure*}[h]
     \centering
     \begin{subfigure}[b]{1.0\textwidth}
         \centering
         \includegraphics[width=\textwidth]{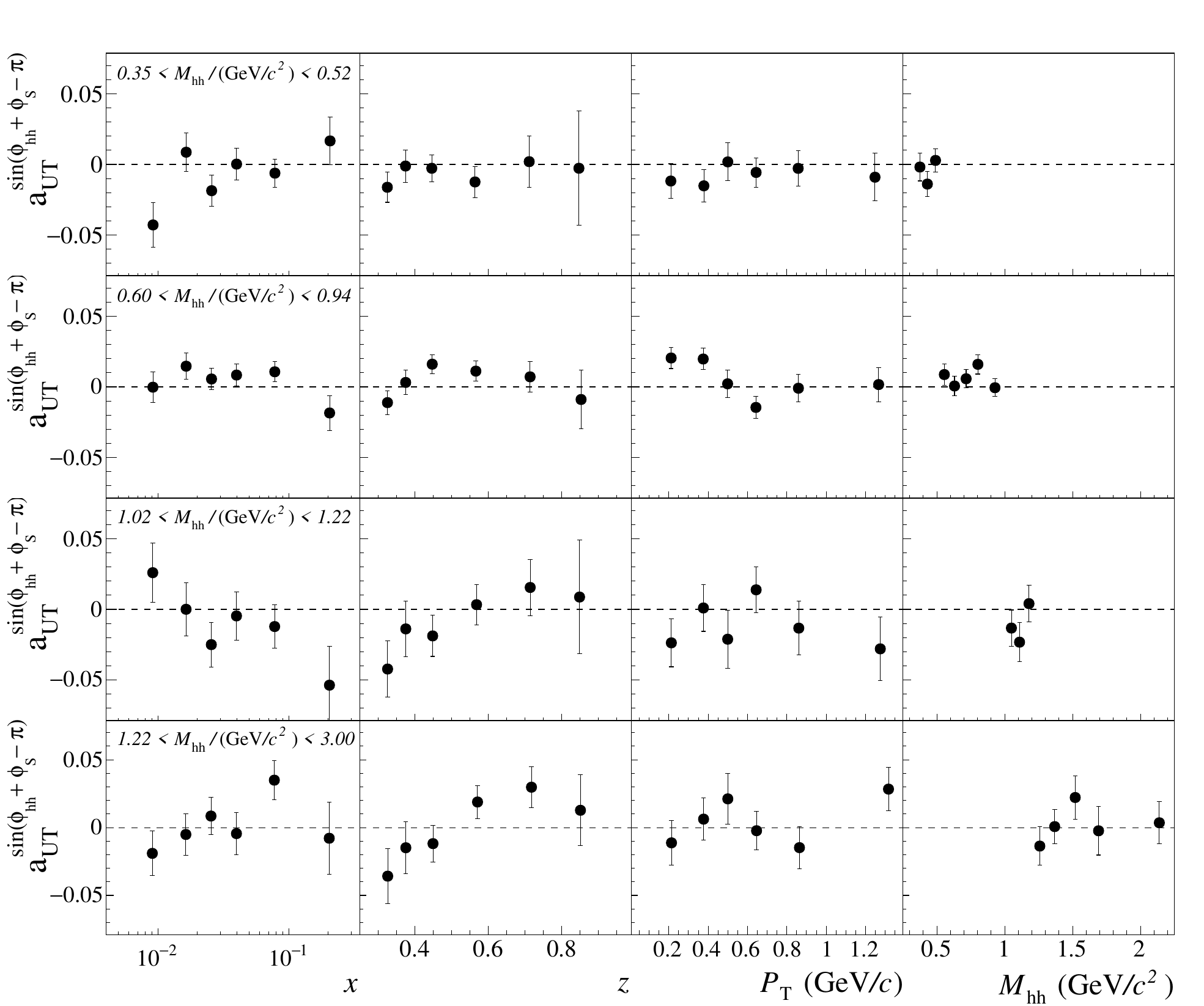}
     \end{subfigure}
        \caption{Collins asymmetry for $h^+h^-$ pairs as a function of the kinematic variables $x$, $z$, $\absptv$ and the invariant mass $\Mhh$ (columns from left to right). The different rows correspond to the invariant mass regions defined in Sec. \ref{sec:procedure}. Only the statistical uncertainties are shown.}
        \label{fig:Collins M}
\end{figure*}

The fraction $f_s$ of \rhoz{} mesons in region II is calculated by dividing the number of \rhoz{} mesons by the total number of $h^+h^-$ pairs in the same region. The contamination from the decay of $f_0$ mesons, estimated to be about $4\%$, is neglected. Also, since the $\rho^0$ distribution is described by a width fixed to the PDG value, possible interference effects with the $\omega(782)$ decays are neglected. The total number of $\rho^0$ mesons is estimated to be $2.6\times 10^6$ and the average signal fraction is $\langle f_s\rangle = 0.18$.

This procedure is applied to all $x$, $z$ and $\absptv$ bins, and the values of $f_s$ are shown in Fig. \ref{fig:sig fraction}. We find $f_s$ to be almost constant and about $0.18$ as a function of $x$ and it increases with $\absptv$ and $z$. The high value (about $0.38$) in the last $z$ bin can be understood in terms of the string fragmentation model, where heavier resonances are produced mostly with large fractional energies \cite{Lund1983}.

As consistency check we compared the counts of $\rho^0$ mesons obtained by summing separately over the $x$, $z$ and $\absptv$ bins, which results in similar values that differ by less than $2\%$ with respect to the integrated value. As a further check, the measured \rhoz{} distribution is compared for each kinematic bin to that expected by using a Breit-Wigner function with mass peak and width fixed to the PDG values and the normalisation fitted to the measured $\rho^0$ distribution in the $\rho^0$ region. The largest differences on the $\rho^0$ counts are found to be less than $7.5\%$ and located in the first two $z$ bins. These differences are taken into account in the evaluation of the systematic uncertainty of the final asymmetries.
Moreover, it is checked that nearly the same background in the $\rho^0$ region can be obtained with an alternative method that combines the invariant mass distributions simulated with the PYTHIA$\,$8 event generator \cite{pythia8} for the different background components (resonant and non-resonant contributions) and Breit-Wigner functions with PDG parameters for the $\rho^0$, $f_0$ and $f_2$ resonances to fit the total $h^+h^-$ distribution up to $\Mhh=1.4\,\rm{GeV}/c^2$. The differences between the two methods are small and are taken into account in the systematic uncertainty of the measured asymmetries for $\rho^0$ mesons.

\section{Results for Collins and Sivers asymmetries}\label{sec:asymmetries}
The Collins asymmetry $\aUTC{}$ for all selected $h^+h^-$ pairs is shown in Fig. \ref{fig:Collins M} as a function of $x$, $z$, $\absptv$ and $\Mhh$. The asymmetries are evaluated in each of the twelve periods of data taking and the final result is obtained as the weighted average. Each row corresponds to a different invariant mass region. Starting from the top, the second row shows the asymmetry in the \rhoz{} region, which has mostly positive values, in particular around $z\sim 0.5$ and for $\absptv<0.5\,\rm{GeV}/c$. This is at variance with the asymmetries in the side band regions, shown in the first and third rows, which tend to be negative. Also, these asymmetries in the side band regions are found to be compatible. No strong kinematic dependence with the invariant mass is found. This is demonstrated by the invariant-mass dependence in the rightmost column. No significant effect is observed at large invariant mass, as can be seen from the bottom row.

\begin{figure}[!h]
     \centering
     \begin{minipage}[b]{0.49\textwidth}
         \centering
         \hspace{-1.5em}
         \includegraphics[width=\textwidth]{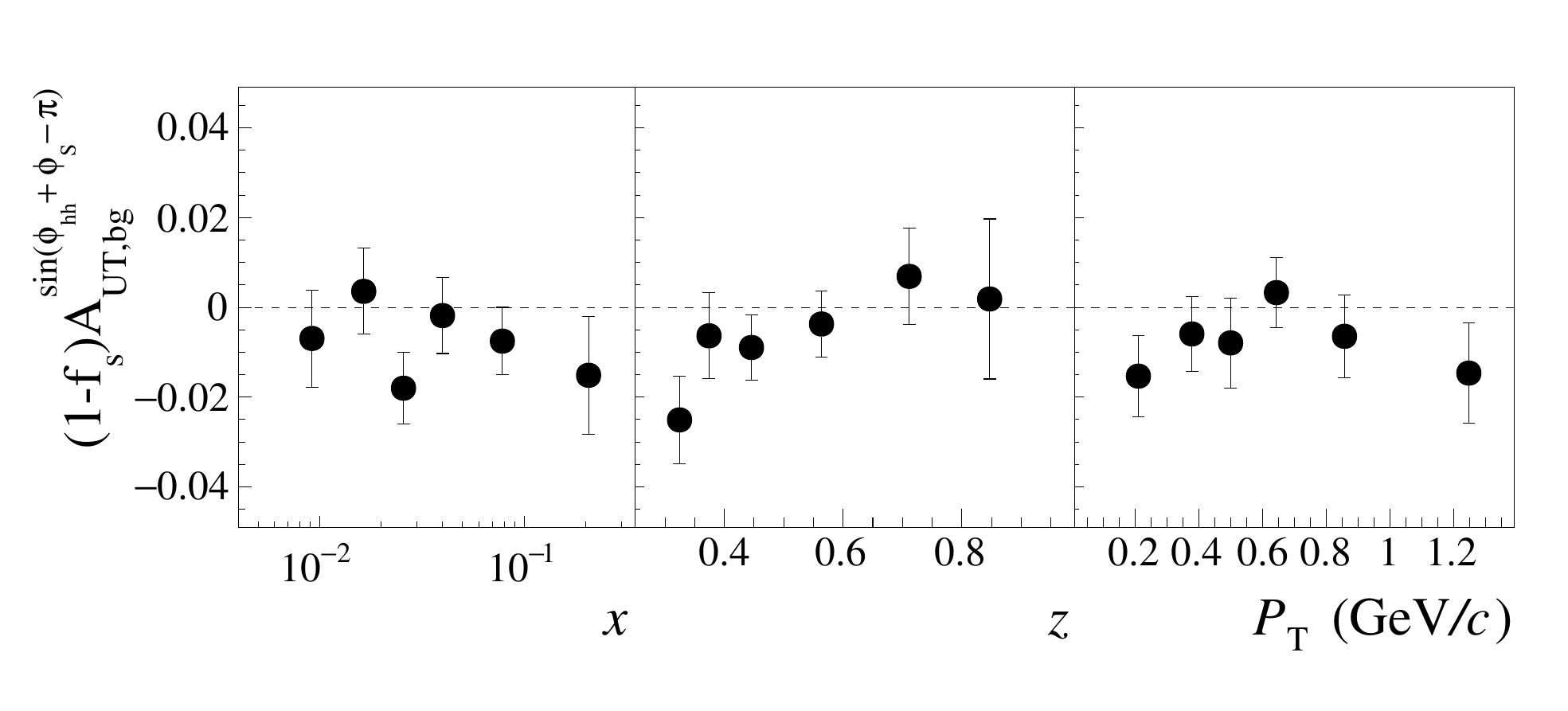}
     \end{minipage}
     \begin{minipage}[b]{0.49\textwidth}
         \centering
         \includegraphics[width=\textwidth]{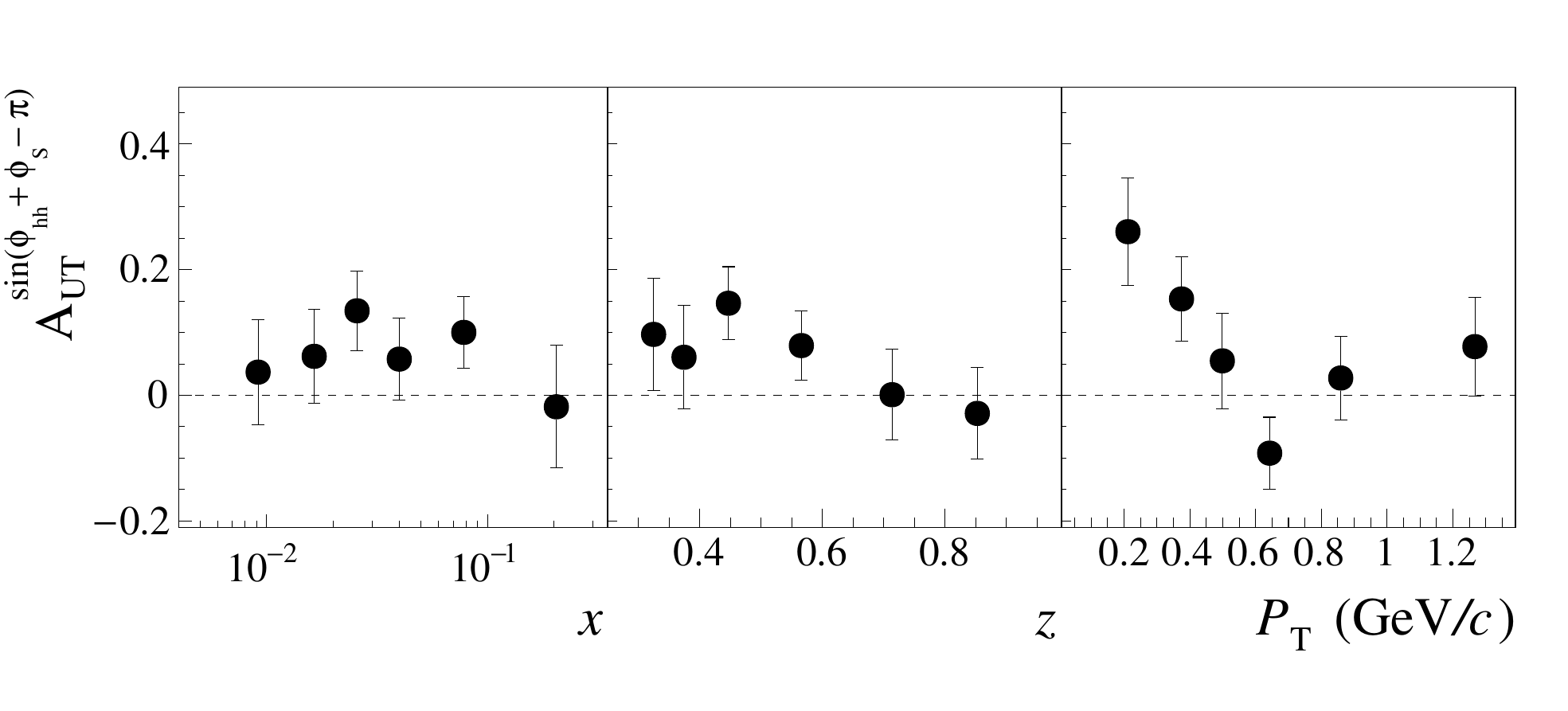}
     \end{minipage}
        \caption{The background contribution to the Collins asymmetry in the \rhoz{} region (top panel) and the final Collins asymmetry for \rhoz{} mesons (bottom panel) as a function of $x$, $z$ and $\absptv$. Only the statistical uncertainties are shown. The systematic uncertainty on the Collins asymmetry for \rhoz{} mesons is estimated to be about $0.6$ the statistical one.}
        \label{fig:Collins signal}
\end{figure}

\begin{figure}[t]
     \centering
     \begin{minipage}[b]{0.5\textwidth}
         \centering
         \includegraphics[width=\textwidth]{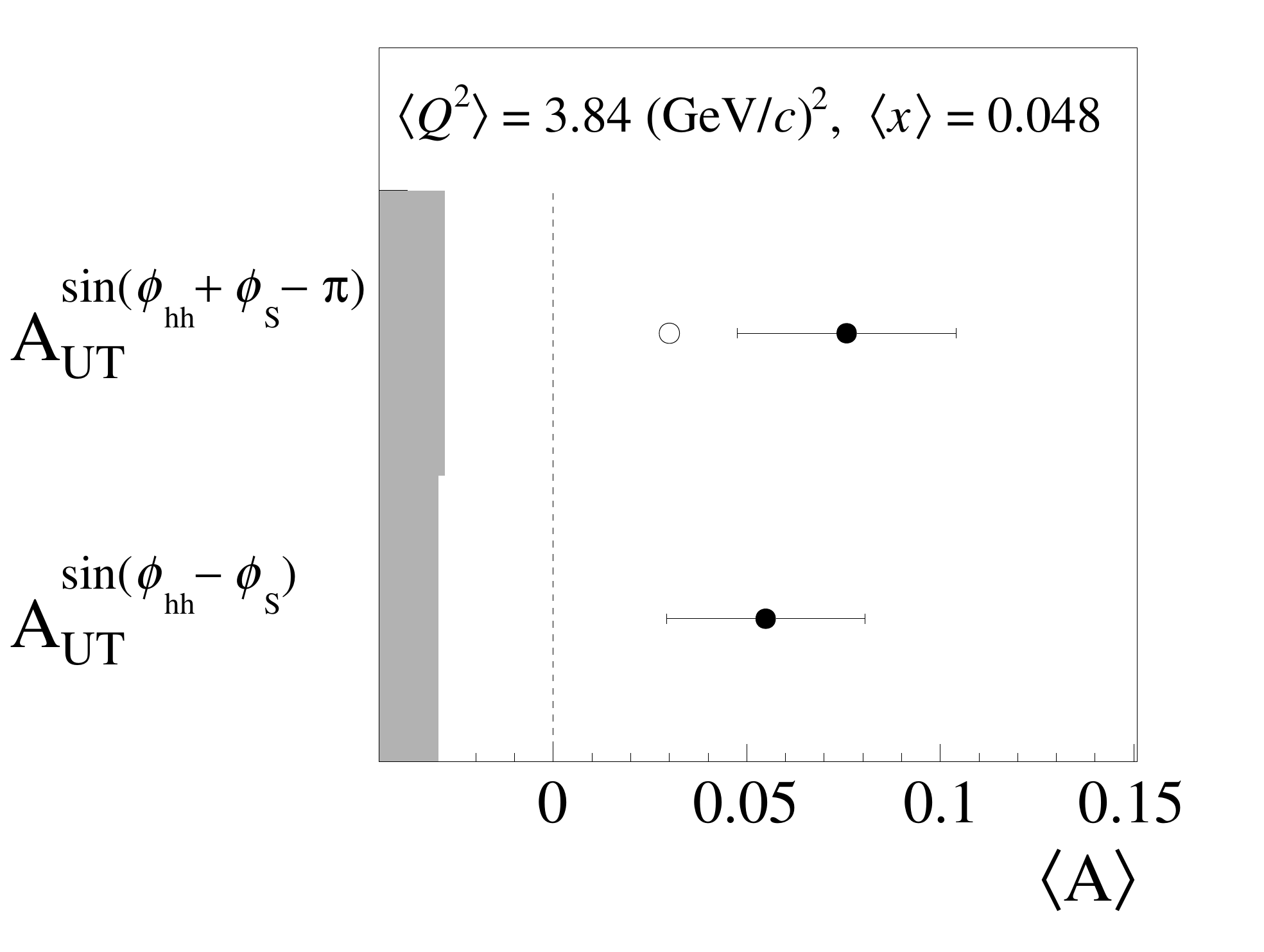}
     \end{minipage}
        \caption{Average values of the Collins and Sivers asymmetries for \rhoz{} mesons. The gray bands represent the evaluated systematic uncertainty. The open point shows the average Collins asymmetry from simulations.}
        \label{fig:rho0 average}
\end{figure}

\begin{figure*}[tbh]
     \centering
     \begin{subfigure}[b]{1.0\textwidth}
         \centering
         \includegraphics[width=\textwidth]{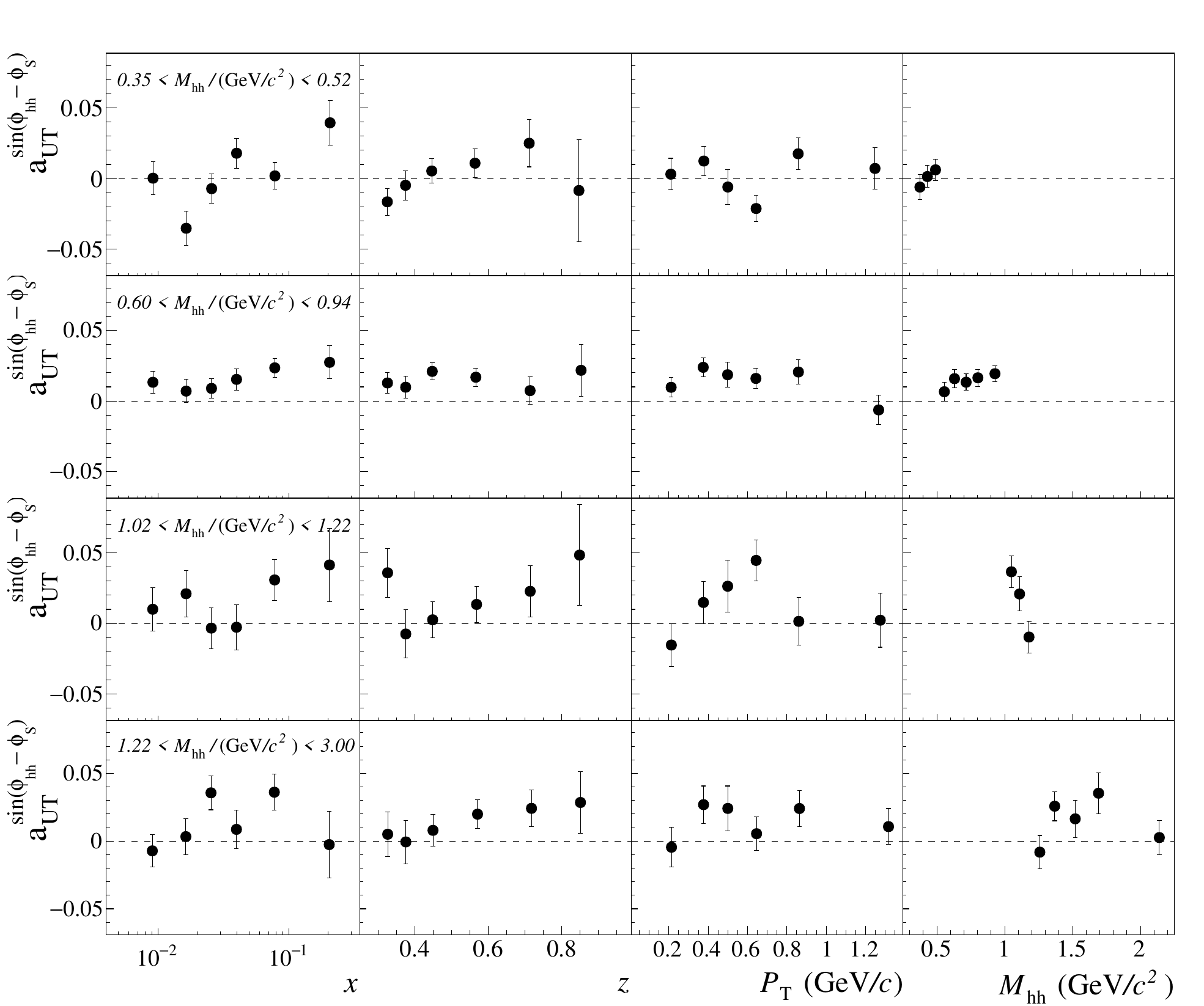}
     \end{subfigure}
        \caption{Sivers asymmetry for $h^+h^-$ pairs as a function of the kinematic variables $x$, $z$, $\absptv$ and the invariant mass $\Mhh$ (columns from left to right). The different rows correspond to the invariant mass regions defined in Sec. \ref{sec:procedure}. Only the statistical uncertainties are shown.}
        \label{fig:Sivers M}
\end{figure*}

\begin{figure}[!h]
     \centering
     \begin{minipage}[b]{0.49\textwidth}
         \centering
          \hspace{-1.5em}
         \includegraphics[width=\textwidth]{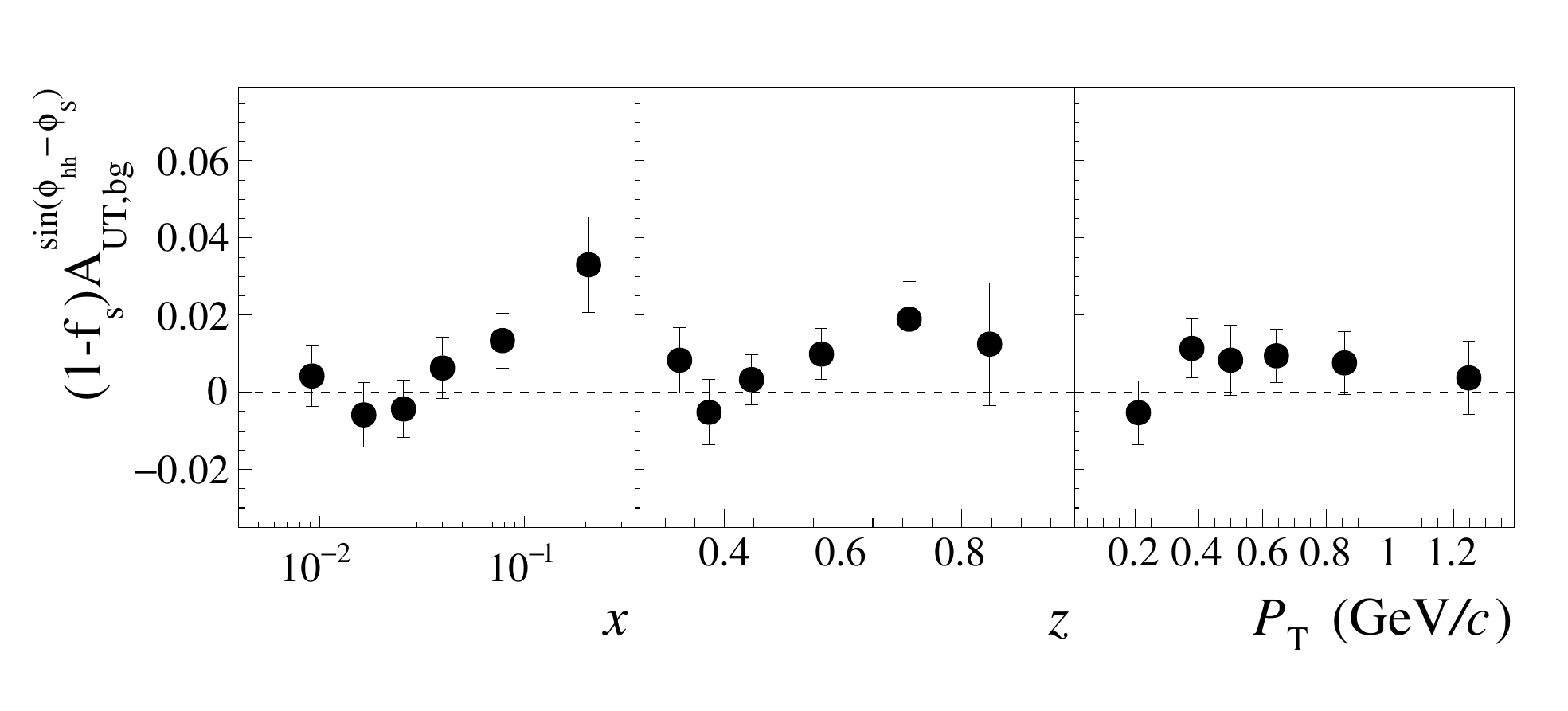}
     \end{minipage}
     \begin{minipage}[b]{0.49\textwidth}
         \centering
         \includegraphics[width=\textwidth]{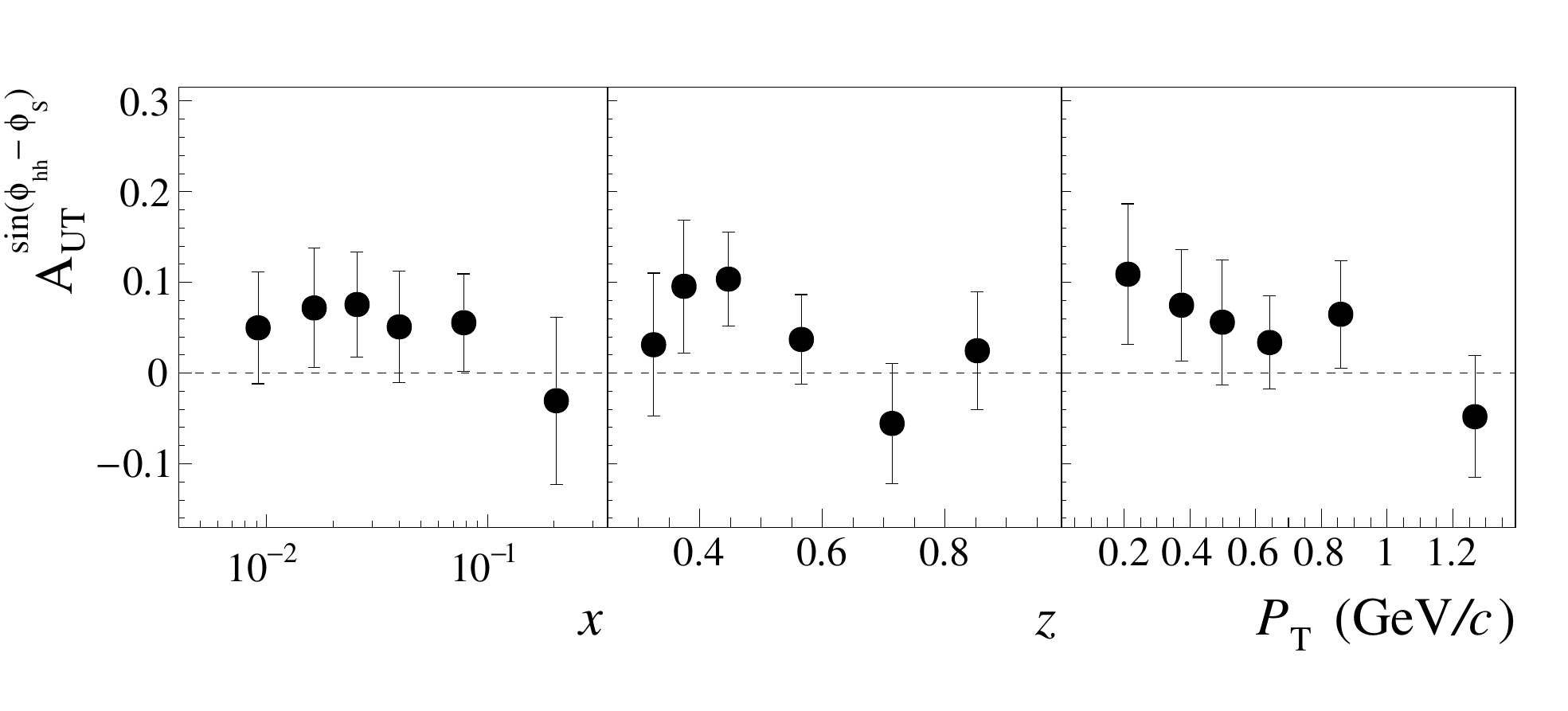}
     \end{minipage}
        \caption{The background contribution to the Sivers asymmetry in the \rhoz{} region (top panel) and the final Sivers asymmetry for \rhoz{} mesons (bottom panel) as a function of $x$, $z$ and $\absptv$. Only the statistical uncertainties are shown. The systematic uncertainty on the Sivers asymmetry for \rhoz{} mesons is estimated to be about $0.6$ the statistical one.}
        \label{fig:Sivers signal}
\end{figure}

\begin{table*}[tb]
\centering
\begin{tabular}{ |p{2.55cm}|p{0.8cm}|p{0.8cm}|p{0.6cm}|p{1.8cm}|p{2.2cm}|p{2.4cm}|p{2.4cm}|  }
 \hline
 $x$-bin \vspace{0.2em}& $\langle x\rangle$ & $\langle y\rangle$ & $\langle z\rangle$ & $\langle \absptv\rangle$ & $\langle Q^2\rangle$ & $A_{\rm{UT}}^{\sin\phiC}$ & $A_{\rm{UT}}^{\sin\phiSiv}$\\
 \hline
 $[0.003,\, 0.013)$ & $0.009$& $0.55$& $0.47$ &$0.60\,\rm{GeV}/c$ &$1.44\,(\rm{GeV}/c)^2$   &$0.037\pm 0.084$ &$0.050\pm 0.062$  \\
 $[0.013,\, 0.020)$ & $0.016$& $0.37$& $0.48$ &$0.58\,\rm{GeV}/c$ &$1.81\,(\rm{GeV}/c)^2$   &$0.062\pm 0.075$ &$0.072\pm 0.066$  \\
 $[0.020,\, 0.032)$ & $0.026$& $0.28$& $0.48$ &$0.56\,\rm{GeV}/c$ &$2.16\,(\rm{GeV}/c)^2$   &$0.134\pm 0.063$ &$0.076\pm 0.058$  \\
 $[0.032,\, 0.050)$ & $0.040$& $0.24$& $0.48$ &$0.54\,\rm{GeV}/c$ &$2.89\,(\rm{GeV}/c)^2$   &$0.058\pm 0.065$ &$0.051\pm 0.062$  \\
 $[0.050,\, 0.130)$ & $0.078$& $0.22$& $0.48$ &$0.54\,\rm{GeV}/c$ &$5.49\,(\rm{GeV}/c)^2$   &$0.100\pm 0.057$ &$0.056\pm 0.054$  \\
 $[0.130,\, 0.700]$ & $0.205$& $0.21$& $0.48$ &$0.56\,\rm{GeV}/c$ &$14.84\,(\rm{GeV}/c)^2$   &$-0.018\pm 0.097$&$-0.030\pm 0.092$ \\
 \hline
 $z$-bin\vspace{0.1em} &$\langle x \rangle$& $\langle y\rangle$ &   $\langle z\rangle$ & $\langle \absptv\rangle$& $\langle Q^2\rangle$ & $A_{\rm{UT}}^{\sin\phiC}$& $A_{\rm{UT}}^{\sin\phiSiv}$\\
 \hline
 $[0.30,\, 0.35)$ &$0.046$ &$0.33$ &$0.32$ &$0.53\,\rm{GeV}/c$ &$3.84\,(\rm{GeV}/c)^2$   &$0.097\pm 0.090$ &$0.031\pm 0.079$  \\
 $[0.35,\, 0.40)$ &$0.048$ &$0.32$ &$0.37$ &$0.54\,\rm{GeV}/c$ &$3.85\,(\rm{GeV}/c)^2$   &$0.061\pm 0.082$ &$0.096\pm 0.073$  \\
 $[0.40,\, 0.50)$ &$0.050$ &$0.31$ &$0.45$ &$0.56\,\rm{GeV}/c$ &$3.86\,(\rm{GeV}/c)^2$   &$0.146\pm 0.058$ &$0.104\pm 0.052$  \\
 $[0.50,\, 0.65)$ &$0.052$ &$0.30$ &$0.57$ &$0.58\,\rm{GeV}/c$ &$3.83\,(\rm{GeV}/c)^2$   &$0.079\pm 0.054$ &$0.037\pm 0.049$  \\
 $[0.65,\, 0.80)$ &$0.052$ &$0.29$ &$0.71$ &$0.59\,\rm{GeV}/c$ &$3.70\,(\rm{GeV}/c)^2$   &$0.001\pm 0.072$ &$-0.055\pm 0.066$ \\
 $[0.80,\, 0.95]$ &$0.038$ &$0.34$ &$0.85$ &$0.54\,\rm{GeV}/c$ &$3.33\,(\rm{GeV}/c)^2$   &$-0.029\pm 0.073$&$0.025\pm 0.065$  \\
 \hline
 $\absptv$-bin\vspace{0.1em} $(\frac{\rm{GeV}}{c})$ & $\langle x\rangle$ & $\langle y\rangle$ & $\langle z\rangle$ & $\langle \absptv\rangle$ & $\langle Q^2\rangle$ & $ A_{\rm{UT}}^{\sin\phiC}$& $A_{\rm{UT}}^{\sin\phiSiv}$\\
 \hline
 $[0.10,\, 0.30)$ &$0.050$ & $0.30$ &$0.47$ &$0.21\,\rm{GeV}/c$  &$3.58\,(\rm{GeV}/c)^2$   &$0.260\pm 0.085$ &$0.109\pm 0.078$  \\
 $[0.30,\, 0.45)$ &$0.050$ & $0.30$ &$0.48$ &$0.37\,\rm{GeV}/c$  &$3.65\,(\rm{GeV}/c)^2$   &$0.153\pm 0.068$ &$0.075\pm 0.061$  \\
 $[0.45,\, 0.55)$ &$0.050$ & $0.30$ &$0.48$ &$0.50\,\rm{GeV}/c$  &$3.73\,(\rm{GeV}/c)^2$   &$0.055\pm 0.076$ &$0.056\pm 0.069$  \\
 $[0.55,\, 0.75)$ &$0.050$ & $0.31$ &$0.48$ &$0.64\,\rm{GeV}/c$  &$3.84\,(\rm{GeV}/c)^2$   &$-0.092\pm 0.057$&$0.034\pm 0.051$  \\
 $[0.75,\, 1.00)$ &$0.049$ & $0.33$ &$0.49$ &$0.86\,\rm{GeV}/c$  &$4.04\,(\rm{GeV}/c)^2$   &$0.028\pm 0.067$ &$0.065\pm 0.059$  \\
 $[1.00,\, 4.00]$ &$0.047$ & $0.38$ &$0.50$ &$1.27\,\rm{GeV}/c$  &$4.49\,(\rm{GeV}/c)^2$   &$0.077\pm 0.078$ &$-0.048\pm 0.067$ \\
 \hline
\end{tabular}
\caption{The measured values of the Collins and Sivers asymmetries for \rhoz{} mesons as a function of $x$, $z$ and $\absptv$. Shown are also the definitions of the $x$, $z$ and $\absptv$ bins, as well as the average values $\langle x\rangle$, $\langle y\rangle$, $\langle z\rangle$, $\langle \absptv\rangle$ and $\langle Q^2\rangle$ for each kinematic bin. Only the statistical uncertainties of the asymmetries are given. The systematic uncertainty is estimated to be about $0.6$ the statistical one.} \label{table:values of rho0 tsa}
\end{table*}

The background asymmetry is evaluated taking the average of the asymmetries in regions I and III. According to Eq.~(\ref{eq:subtraction}), the background asymmetry is first rescaled by the factor $1-f_s$ and then subtracted from the Collins asymmetry in the \rhoz{} region. The contribution of the background asymmetry in the \rhoz{} region is shown in the top panel of Fig. \ref{fig:Collins signal} as a function of $x$, $z$ and $\absptv$. It has mostly negative values, as expected from Fig. \ref{fig:Collins M}, although the uncertainties are large. The final Collins asymmetry for \rhoz{} mesons is shown in the bottom panel of Fig. \ref{fig:Collins signal} and the corresponding values are given in Tab. \ref{table:values of rho0 tsa}.
Given the large uncertainties, no clear trend can be seen as a function of $x$ and $z$. As a function of $\absptv$ the asymmetry increases for $\absptv<0.5\, \gevcc\,$, as suggested by the simulations carried out with the recursive string+${}^3P_0$ model in Ref. \cite{Kerbizi:2021}, using the scenario where the production of vector mesons with longitudinal polarization in the GNS is favoured.

The systematic uncertainty on the extracted transverse single spin asymmetries for \rhoz{} mesons is estimated to be about $0.6$ times the statistical one. As described above, this estimate accounts for the systematic uncertainty on the evaluation of the background distribution in the \rhoz{} region. In addition it takes into account other sources of systematic uncertainties, such as the period by period compatibility of the asymmetries and variations on the \rhoz{} asymmetries induced by changing the invariant mass interval corresponding to the region II.

The average value of the Collins asymmetry for \rhoz{} mesons is shown in Fig. \ref{fig:rho0 average}. The asymmetry is positive with a significance of $2.3$ standard deviations, evaluated taking into account both statistical and systematic uncertainties, and is in agreement with the model predictions \cite{Czyzewski-vm}. Comparing the average value of the measured \rhoz{} Collins asymmetry with that of the simulated one \cite{Kerbizi:2021}, shown as the open point in Fig. \ref{fig:rho0 average}, consistency is found within about one standard deviation.

The same procedure is repeated for the Sivers asymmetry. The asymmetry $\aUTS{}$ is shown as a function of $x$, $z$, $\absptv$ and $\Mhh$ in Fig. \ref{fig:Sivers M} for the different invariant mass regions. The Sivers asymmetry in the \rhoz{} region exhibits positive values and a clear trend with $x$. Contrary to the case of the Collins asymmetry, the Sivers asymmetry is positive and significant also in the side-bands. This indicates that the contribution of the background to the asymmetry is large in the \rhoz{} region, as shown in the top panel of Fig. \ref{fig:Sivers signal}. The background-subtracted final Sivers asymmetry for \rhoz{} mesons is shown in the bottom panel in Fig. \ref{fig:Sivers signal} and the corresponding values are given in Tab. \ref{table:values of rho0 tsa}. The shown uncertainties are the statistical ones. As for the Collins asymmetry, the systematic uncertainty is evaluated to be $0.6$ times the statistical one. The average value of the asymmetry is shown in Fig. \ref{fig:rho0 average}. It is found to be positive with a significance of $1.8$ standard deviations. A positive Sivers asymmetry for \rhoz{} mesons is expected because, by momentum conservation in the hard scattering, the Sivers function induces a modulation on the direction of the struck quark which propagates to all the hadrons produced in the fragmentation process. The Sivers asymmetry for \rhoz{} mesons is thus naively expected to be similar to the average value of the Sivers asymmetries for positive and negative pions, which is positive \cite{COMPASS-collins-sivers}. Given the large uncertainties, no clear trends as a function of the kinematic variables can be seen for the \rhoz{} asymmetry.

\section{Conclusions}\label{sec:conclusions}
The COMPASS Collaboration has performed the first measurement of the Collins and Sivers transverse single spin asymmetries for \rhoz{} mesons produced in DIS off transversely polarized protons. The full data set of SIDIS events collected by COMPASS in 2010 was analysed. An indication for a positive Collins asymmetry is found. The result is in agreement with the expectation from the recursive string+${}^3P_0$ model of the polarized quark fragmentation process. Also an indication for a positive Sivers asymmetry is found, in agreement with the parton model. These measurements are complementary to the single hadron asymmetries, since they give new input to studying and understanding of the hadronisation process. This work shows that the measurement of TSAs for inclusive vector meson production in DIS is feasible and could be done with higher precision at future facilities.

\section*{Acknowledgements}
This work was made possible thanks to the financial support of our funding agencies. We also acknowledge the support of the CERN management and staff, as well as the skills and efforts of the technicians of the collaborating institutes.

\bibliographystyle{utphys}

\clearpage
\textbf{The COMPASS Collaboration}

\vspace{10pt}
\begin{flushleft}
G.~D.~Alexeev$^\textrm{{\footnotesize\hyperlink{hl:dubna}{29}}}$\orcidlink{0009-0007-0196-8178},
M.~G.~Alexeev$^\textrm{{\footnotesize\hyperlink{hl:turin_u}{20},\hyperlink{hl:turin_i}{19}}}$\orcidlink{0000-0002-7306-8255},
C.~Alice$^\textrm{{\footnotesize\hyperlink{hl:turin_u}{20},\hyperlink{hl:turin_i}{19}}}$\orcidlink{0000-0001-6297-9857},
A.~Amoroso$^\textrm{{\footnotesize\hyperlink{hl:turin_u}{20},\hyperlink{hl:turin_i}{19}}}$\orcidlink{0000-0002-3095-8610},
V.~Andrieux$^\textrm{{\footnotesize\hyperlink{hl:illinois}{33}}}$\orcidlink{0000-0001-9957-9910},
V.~Anosov$^\textrm{{\footnotesize\hyperlink{hl:dubna}{29}}}$\orcidlink{0009-0003-3595-9561},
K.~Augsten$^\textrm{{\footnotesize\hyperlink{hl:praguectu}{4}}}$\orcidlink{0000-0001-8324-0576},
W.~Augustyniak$^\textrm{{\footnotesize\hyperlink{hl:warsaw}{24}}}$,
C.~D.~R.~Azevedo$^\textrm{{\footnotesize\hyperlink{hl:aveiro}{27}}}$\orcidlink{0000-0002-0012-9918},
B.~Badelek$^\textrm{{\footnotesize\hyperlink{hl:warsawu}{26}}}$\orcidlink{0000-0002-4082-1466}
% ...
\hypertarget{hl:calcutta}{$^\textrm{{\footnotesize 14}}$\footnotesize~Matrivani Institute of Experimental Research \& Education, Calcutta-700 030, India$^\textrm{{\tiny\hyperlink{hl:C}{C}}}$\\}
\hypertarget{hl:telaviv}{$^\textrm{{\footnotesize 15}}$\footnotesize~Tel Aviv University, School of Physics and Astronomy, 69978 Tel Aviv, Israel$^\textrm{{\tiny\hyperlink{hl:D}{D}}}$\\}
\hypertarget{hl:triest_a}{$^\textrm{{\footnotesize 16}}$\footnotesize~Abdus Salam ICTP, 34151 Trieste, Italy\\}
\hypertarget{hl:triest_i}{$^\textrm{{\footnotesize 17}}$\footnotesize~Trieste Section of INFN, 34127 Trieste, Italy\\}
\hypertarget{hl:triest_u}{$^\textrm{{\footnotesize 18}}$\footnotesize~University of Trieste, Dept.\ of Physics, 34127 Trieste, Italy\\}
\hypertarget{hl:turin_i}{$^\textrm{{\footnotesize 19}}$\footnotesize~Torino Section of INFN, 10125 Torino, Italy\\}
\hypertarget{hl:turin_u}{$^\textrm{{\footnotesize 20}}$\footnotesize~University of Torino, Dept.\ of Physics, 10125 Torino, Italy\\}
\hypertarget{hl:miyazaki}{$^\textrm{{\footnotesize 21}}$\footnotesize~University of Miyazaki, Miyazaki 889-2192, Japan$^\textrm{{\tiny\hyperlink{hl:E}{E}}}$\\}
\hypertarget{hl:nagoya}{$^\textrm{{\footnotesize 22}}$\footnotesize~Nagoya University, 464 Nagoya, Japan$^\textrm{{\tiny\hyperlink{hl:E}{E}}}$\\}
\hypertarget{hl:yamagata}{$^\textrm{{\footnotesize 23}}$\footnotesize~Yamagata University, Yamagata 992-8510, Japan$^\textrm{{\tiny\hyperlink{hl:E}{E}}}$\\}
\hypertarget{hl:warsaw}{$^\textrm{{\footnotesize 24}}$\footnotesize~National Centre for Nuclear Research, 02-093 Warsaw, Poland$^\textrm{{\tiny\hyperlink{hl:F}{F}}}$\\}
\hypertarget{hl:warsawtu}{$^\textrm{{\footnotesize 25}}$\footnotesize~Warsaw University of Technology, Institute of Radioelectronics, 00-665 Warsaw, Poland$^\textrm{{\tiny\hyperlink{hl:F}{F}}}$\\}
\hypertarget{hl:warsawu}{$^\textrm{{\footnotesize 26}}$\footnotesize~University of Warsaw, Faculty of Physics, 02-093 Warsaw, Poland$^\textrm{{\tiny\hyperlink{hl:F}{F}}}$\\}
\hypertarget{hl:aveiro}{$^\textrm{{\footnotesize 27}}$\footnotesize~University of Aveiro, I3N, Dept. of Physics, 3810-193 Aveiro, Portugal$^\textrm{{\tiny\hyperlink{hl:G}{G}}}$\\}
\hypertarget{hl:lisbon}{$^\textrm{{\footnotesize 28}}$\footnotesize~LIP, 1649-003 Lisbon, Portugal$^\textrm{{\tiny\hyperlink{hl:G}{G}}}$\\}
\hypertarget{hl:dubna}{$^\textrm{{\footnotesize 29}}$\footnotesize~Affiliated with an international laboratory covered by a cooperation agreement with CERN\\}
\hypertarget{hl:russia}{$^\textrm{{\footnotesize 30}}$\footnotesize~Affiliated with an institute covered by a cooperation agreement with CERN.\\}
\hypertarget{hl:cern}{$^\textrm{{\footnotesize 31}}$\footnotesize~CERN, 1211 Geneva 23, Switzerland\\}
\hypertarget{hl:taipei}{$^\textrm{{\footnotesize 32}}$\footnotesize~Academia Sinica, Institute of Physics, Taipei 11529, Taiwan$^\textrm{{\tiny\hyperlink{hl:H}{H}}}$\\}
\hypertarget{hl:illinois}{$^\textrm{{\footnotesize 33}}$\footnotesize~University of Illinois at Urbana-Champaign, Dept.\ of Physics, Urbana, IL 61801-3080, USA$^\textrm{{\tiny\hyperlink{hl:I}{I}}}$\\}

\vspace{10pt}
\hypertarget{hl:*}{$^\textrm{{\footnotesize *}}$\footnotesize~Corresponding author\\}
\hypertarget{hl:a}{$^\textrm{{\footnotesize a}}$\footnotesize~Supported by the European Union’s Horizon 2020 research and innovation programme under grant agreement STRONG–2020 - No 824093\\}
\hypertarget{hl:b}{$^\textrm{{\footnotesize b}}$\footnotesize~Supported by ANR, France with P2IO LabEx (ANR-10-LABX-0038)  in the framework "Investissements d'Avenir" (ANR-11-IDEX-0003-01)\\}
\hypertarget{hl:c}{$^\textrm{{\footnotesize c}}$\footnotesize~Supported by the DFG Research Training Group Programmes 1102 and 2044 (Germany)\\}
\hypertarget{hl:d}{$^\textrm{{\footnotesize d}}$\footnotesize~Retired from Ludwig-Maximilians-Universit\"at, 80539 M\"unchen, Germany\\}
\hypertarget{hl:d1}{$^\textrm{{\footnotesize d1}}$\footnotesize~Supported by the DFG cluster of excellence `Origin and Structure of the Universe' (www.universe-cluster.de) (Germany)\\}
\hypertarget{hl:e}{$^\textrm{{\footnotesize e}}$\footnotesize~Also at ORIGINS Excellence Cluster, 85748 Garching, Germany\\}
\hypertarget{hl:f}{$^\textrm{{\footnotesize f}}$\footnotesize~Also at Institut f\"ur Theoretische Physik, Universit\"at T\"ubingen, 72076 T\"ubingen, Germany\\}
\hypertarget{hl:g}{$^\textrm{{\footnotesize g}}$\footnotesize~Present address: NISER, Centre for Medical and Radiation Physics, Bubaneswar, India\\}
\hypertarget{hl:h}{$^\textrm{{\footnotesize h}}$\footnotesize~Also at University of Eastern Piedmont, 15100 Alessandria, Italy\\}
\hypertarget{hl:h1}{$^\textrm{{\footnotesize h1}}$\footnotesize~Supported by the Funds for Research 2019-22 of the University of Eastern Piedmont\\}
\hypertarget{hl:i}{$^\textrm{{\footnotesize i}}$\footnotesize~Also at Chubu University, Kasugai, Aichi 487-8501, Japan\\}
\hypertarget{hl:j}{$^\textrm{{\footnotesize j}}$\footnotesize~Also at KEK, 1-1 Oho, Tsukuba, Ibaraki 305-0801, Japan\\}
\hypertarget{hl:k}{$^\textrm{{\footnotesize k}}$\footnotesize~Also at Dept.\ of Physics, Pusan National University, Busan 609-735, Republic of Korea\\}
\hypertarget{hl:k1}{$^\textrm{{\footnotesize k1}}$\footnotesize~Also at Physics Dept., Brookhaven National Laboratory, Upton, NY 11973, USA\\}
\hypertarget{hl:l}{$^\textrm{{\footnotesize l}}$\footnotesize~Also at Dept.\ of Physics, National Central University, 300 Jhongda Road, Jhongli 32001, Taiwan\\}
\hypertarget{hl:m}{$^\textrm{{\footnotesize m}}$\footnotesize~Also at Dept.\ of Physics, National Kaohsiung Normal University, Kaohsiung County 824, Taiwan\\}
\hypertarget{hl:$\dagger$}{$^\textrm{{\footnotesize $\dagger$}}$\footnotesize~Deceased\\}

\vspace{10pt}
\hypertarget{hl:A}{$^\textrm{{\tiny A}}$\footnotesize~Supported by MEYS, Grants LM2015058, LM2018104 and LTT17018 and Charles University Grant PRIMUS/22/SCI/017 (Czech Republic)\\}
\hypertarget{hl:B}{$^\textrm{{\tiny B}}$\footnotesize~Supported by BMBF - Bundesministerium f\"ur Bildung und Forschung (Germany)\\}
\hypertarget{hl:C}{$^\textrm{{\tiny C}}$\footnotesize~Supported by B. Sen fund (India)\\}
\hypertarget{hl:D}{$^\textrm{{\tiny D}}$\footnotesize~Supported by the Israel Academy of Sciences and Humanities (Israel)\\}
\hypertarget{hl:E}{$^\textrm{{\tiny E}}$\footnotesize~Supported by MEXT and JSPS, Grants 18002006, 20540299, 18540281 and 26247032, the Daiko and Yamada Foundations (Japan)\\}
\hypertarget{hl:F}{$^\textrm{{\tiny F}}$\footnotesize~Supported by NCN, Grant 2020/37/B/ST2/01547 (Poland)\\}
\hypertarget{hl:G}{$^\textrm{{\tiny G}}$\footnotesize~Supported by FCT, Grants CERN/FIS-PAR/0022/2019 and CERN/FIS-PAR/0016/2021 (Portugal)\\}
\hypertarget{hl:H}{$^\textrm{{\tiny H}}$\footnotesize~Supported by the Ministry of Science and Technology (Taiwan)\\}
\hypertarget{hl:I}{$^\textrm{{\tiny I}}$\footnotesize~Supported by the National Science Foundation, Grant no. PHY-1506416 (USA)\\}

\end{flushleft}

\end{document}